%% file: main.tex
\newif\ifSingleColumn
\SingleColumnfalse
\ifSingleColumn
\documentclass[draftclsnofoot, onecolumn, 12pt]{IEEEtran}
\else
\documentclass[journal]{IEEEtran}
\fi
\usepackage[T1]{fontenc}
\usepackage{amsmath,amssymb,amsfonts}
\usepackage{array}
\usepackage{graphicx}
\usepackage{booktabs}
\usepackage{multirow}
\usepackage{float}
\usepackage{bm}
\usepackage{xspace}
\usepackage{enumitem}
\usepackage{cite}
\usepackage{algorithm}
\usepackage[noend]{algpseudocode}
\usepackage{xcolor}
\usepackage{url}
\usepackage{comment}
\usepackage[hidelinks]{hyperref}

\input{defines}

\begin{document}

\title{Token Encoding for Semantic Recovery}

\author{Jingzhi~Hu,~\IEEEmembership{Member,~IEEE,}
Ouya~Wang,~\IEEEmembership{Member,~IEEE,}
Geoffrey~Ye~Li,~\IEEEmembership{Fellow,~IEEE}
\thanks{A preliminary version of this work was accepted for presentation at the 2026 IEEE GLOBECOM~\cite{Hu26GLOBECOM_TokCode}.}
\thanks{J. Hu, O. Wang, and G. Y. Li are with the Department of Electrical and Electronic Engineering, Imperial College London, London SW7 2AZ, UK. Email: jingzhi.hu518@gmail.com, ouya.wang20@imperial.ac.uk, geoffrey.li@imperial.ac.uk.}}


\maketitle

\input{sections/1_abstract}
\ifSingleColumn
\newpage
\fi
\input{sections/2_introduction}
\input{sections/4_preliminaries}
\input{sections/5_token_encoding_framework}
\input{sections/6_cadet}
\input{sections/6a_stage1_expert_tuning}
\input{sections/6b_stage2_distillation}
\input{sections/6c_stage3_rl_refinement}

\input{sections/7a_experiments_setup}

\input{sections/7b_experiments_results}
\input{sections/9_conclusion}

\appendices

\bibliographystyle{IEEEtran}
\bibliography{biblio}

\end{document}

%% file: defines.tex
\newcommand{\sys}{\textsc{TokCode}\xspace}
\newcommand{\alg}{\textsc{Cadet}\xspace}

\newcommand{\bJ}{{\bm J}}

\newcommand{\cV}{{\mathcal{V}}}

\newcommand{\cR}{{\mathcal{R}}}

\newcommand{\cL}{{\mathcal{L}}}
\newcommand{\cD}{{\mathcal{D}}}

\newcommand{\cT}{{\mathcal{T}}}

\newcommand{\bR}{{\mathbb{R}}}

\newcommand{\beq}{\begin{equation}}
\newcommand{\eeq}{\end{equation}}

\newcommand{\figwidth}{0.95\linewidth}

%% file: sections/1_abstract.tex
\begin{abstract}
In generative semantic communication, semantic tokens guide receiver-side generative models to synthesize high-dimensional content.
In challenging network environments, however, frequent token erasure distorts the conveyed semantics beyond what receiver-side recovery can restore.
In this paper, we propose a token encoding framework (\sys) for robust semantic recovery, achieving erasure resilience by restructuring redundancy in the semantic domain.
\sys uses a lightweight adapter to recast a general-purpose large language model (LLM) at the transmitter into a token encoder, exploiting the LLM's pretrained semantic prior to avoid introducing a dedicated deep model.
To optimize the adapter efficiently and make it applicable across diverse channels, we develop a channel-quality-aware distillation approach for token encoder training~(\alg).
Using a differentiable sentence-level semantic surrogate, \alg tunes T5 foundation models into experts for distinct erasure rates and distills them into a single reconfigurable low-rank adapter, enabling subsequent reinforcement learning (RL) to start above the plateau where direct RL stalls.
Simulation results on token-based generative image transmission show that \sys improves the image-level similarity over the best-performing receiver-side recovery benchmark by $14.1\%$--$22.4\%$, closing $71.9\%$--$76.5\%$ of its gap to the erasure-aware oracle encoding, when only $20\%$ to $50\%$ of the tokens survive.
\end{abstract}

\begin{IEEEkeywords}
Token encoding, generative semantic communication, semantic recovery, challenging network environments.
\end{IEEEkeywords}

%% file: sections/2_introduction.tex
\section{Introduction}
\label{sec_intro}

Future wireless networks are envisioned to be ubiquitous, expanding coverage into \emph{challenging environments}, such as remote, offshore, underwater, and underground scenarios~\cite{Imran24PIEEE_Exploring,Tataria21PIEEE_6G,Jouhari19ACCESS_Underwater,Hasan25ACCESS_Underground}.
Enabling communication in these challenging wireless environments is critical for essential operations, such as natural resource exploration, environmental monitoring, infrastructure maintenance, scientific expeditions, and emergency response~\cite{Saeed19COMST_IoUT,Jahanbakht21COMST_IoUWT,Pervez18ACCESS_Emergency}.
In such environments, scarce channel capacity often limits conventional communication systems to low-rate traffic, such as text messages, sensor readings, and status parameters.
High-dimensional data, such as images and videos, can support more intuitive and immersive user experiences, however, remain challenging to deliver.

Generative semantic communication has emerged as a promising paradigm for such environments~\cite{Liang25TCCN_GAISemCom,Jiang25JSAC_FMSAT}.
Rather than transmitting the raw high-dimensional data, it allows the transmitter to send compact \emph{semantic information}, from which a receiver-side generative model synthesizes the target content.
As the wireless link only needs to carry the compact semantic information, generative semantic communication systems require significantly less bandwidth than conventional communication systems.

A representative form of semantic information is the textual description of the high-dimensional data~\cite{Ren25TVT_GenSemComPrompt}.
As text is the native input modality of generative models, a textual description can drive content generation at the receiver without any intermediate decoding stage.
Moreover, compared with semantic feature embeddings, such a description is human-interpretable and usually more compact.
Tokenization techniques compress such a description even further by segmenting the text into pieces and representing each piece by its index in a vocabulary, so that the semantic information is finally carried by a token sequence with a very high compression ratio.

However, communication with token sequences remains fragile, especially in challenging wireless environments.
Severe fading, unpredictable interference, and propagation delays all lead to a non-negligible residual packet-loss rate even under the most robust modulation and forward error correction.
As a compact token sequence inherently carries little redundancy, and any of its tokens may carry information critical to the overall meaning, the loss of tokens may cause severe semantic distortion, leading the receiver to generate content that no longer reflects the intended meaning.

To handle this issue, existing methods exploit semantic correlation among tokens to combat channel corruption, mainly relying on jointly trained transceivers or receiver-side restoration.
In~\cite{Hu23TWC_Robust}, a vector-quantized variational autoencoder (VQ-VAE) is trained end-to-end to reconstruct randomly masked image tokens, improving robustness to noise while reducing the transmitted payloads.
In~\cite{Peng24TWC_RobustText}, R-DeepSC estimates per-token corruption probabilities and downweights unreliable tokens through calibrated self-attention in a jointly trained semantic transceiver, thereby limiting their impact on the reconstructed sentence.
In~\cite{Qiao25WCM_Token}, a large neural model is exploited at the receiver to process token sequences and recover missing tokens under packet loss.
In~\cite{Liu25SPAWC_ResiTok}, images are encoded into hierarchical key and detail tokens, and the decoder is trained to reconstruct them when a randomly selected suffix of detail tokens is replaced with zeros.
In~\cite{Han26TMC_Glaris}, speech is encoded as a VQ-VAE token sequence and transformed into packetized codes; upon packet loss, the receiver reconstructs the missing token representations using learned sequence dependencies and auxiliary redundancy.

In this paper, we focus on transmitter-side token encoding for robust semantic recovery without costly end-to-end training.
The following example motivates our work.
If the sentence, ``{a cartoon rabbit riding a red bicycle on the grass},'' loses either ``{rabbit},'' ``{red},'' or ``{grass},'' its original meaning will be hard to infer, as many equally plausible choices may fit the remaining context.
Therefore, before channel transmission, the token sequence should be proactively \emph{encoded} into a more robust one at the semantic level.
Moreover, we require the encoded token sequence to use the same vocabulary as the source and to occupy no more tokens than it, so that no additional vocabulary is needed and the bandwidth efficiency of token-based transmission is preserved.

Nevertheless, such an encoding is highly challenging.
As we allow no extra tokens to be added, robustness can be gained only by substituting tokens within the source sequence.
The encoding must therefore recognize which tokens hold critical semantic information and should be left unchanged, and which tokens can be substituted to make the conveyed meaning more robust against random channel erasure.
The encoding is further complicated by the structural roles of tokens since a token that appears insignificant individually may determine how the tokens relate to each other.
Satisfying these requirements demands an accurate understanding of semantic information at every level, from individual tokens to their interdependencies and to the sequence as a whole.

\ifSingleColumn
    \renewcommand{\figwidth}{0.7\linewidth}
\else
    \renewcommand{\figwidth}{1\linewidth}
\fi

\begin{figure}[t]
  \centering
  \includegraphics[width=\figwidth]{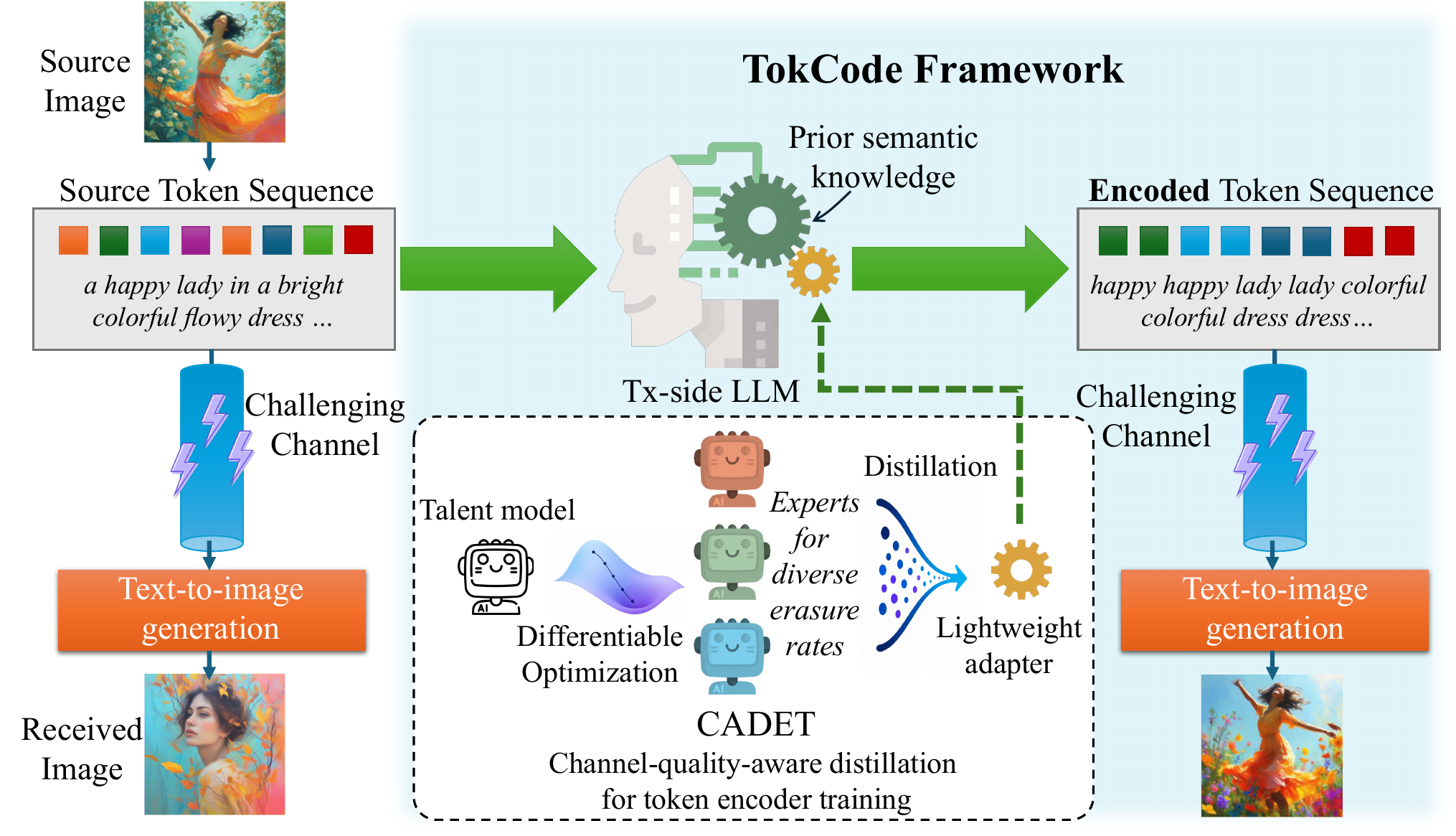}
  \ifSingleColumn
  \else
  \vspace{-1em}
  \fi
  \caption{Working principle of the \sys framework.}
  \ifSingleColumn
  \else
  \vspace{-1em}
  \fi
  \label{fig_intro_illustration}
\end{figure}

As shown in Fig.~\ref{fig_intro_illustration}, we propose a token encoding framework~(\sys) for accurate semantic recovery in this paper.
We exploit the fact that the semantic understanding demanded by the token encoding is precisely what a pretrained large language model~(LLM) intrinsically possesses, and that the growing deployment of local LLMs makes such a model available at the transmitter.
\sys therefore recasts a general-purpose local LLM into a token encoder through a lightweight pluggable adapter, which preserves the model's prior semantic understanding while redirecting it toward producing loss-resilient token sequences.

Training an adapter of an LLM relies on supervised fine-tuning on a large annotated dataset, or on reinforcement learning against a downstream task reward.
However, loss-resilient token encoding has no ground truth, and its reward can be obtained only from the downstream generated content, which is expensive to evaluate and varies with the channel conditions, leaving reinforcement learning alone costly and without a reliable learning signal to guide the search.
To optimize the adapter efficiently and make it applicable across diverse channel conditions, we develop a \textbf{c}hannel-quality-\textbf{a}ware \textbf{d}istillation approach for token \textbf{e}ncoder \textbf{t}raining~(\alg).
Specifically, \alg first replaces the costly image-level reward with a differentiable sentence-level semantic surrogate.
Then, it forms erasure-rate-specific token-encoding experts by directly optimizing the differentiable surrogate over \emph{talent models}, whose architectures and pretraining are better aligned with token encoding than those of general-purpose LLMs.
The encoding strategies of these experts are then distilled into a single reconfigurable low-rank adapter covering the full range of channel conditions, so that the subsequent reinforcement learning departs from a point already beyond the plateau at which direct reinforcement learning stalls.

We take a generative semantic communication system for the typical image transmission task as a case study, and implement both \sys and \alg on the open-source Qwen LLM~\cite{Yang25arXiv_Qwen3}.
Extensive experiments are conducted on a large set of real textual prompts under a range of channel conditions, comparing \sys against baselines that recover the semantics with receiver-side LLMs and against an oracle with \emph{a posteriori} knowledge of the erasure pattern.
Performance is assessed both by the sentence-level semantic similarity between the received and source sequences and by the similarity between the images generated from them.
The main contributions of this paper are summarized as follows.
\begin{itemize}[leftmargin=*]
\item We formulate the token encoding problem for semantic recovery and propose the \sys framework to solve it, recasting a general-purpose local LLM into a token encoder through a lightweight pluggable adapter.

\item Our \alg overcomes the absence of encoding ground truth and the inefficiency of using image-level rewards, expediting the adapter training through distillation to avoid the plateau of direct reinforcement learning.

\item Based on our simulation results, when only $20\%$ to $50\%$ of the tokens survive, \sys closes $71.9\%$--$76.5\%$ of the gap between the best receiver-side recovery benchmark and the erasure-aware oracle encoding.
\end{itemize}

Compared with our preliminary conference version~\cite{Hu26GLOBECOM_TokCode}, this paper extends the framework to a more practical scenario, where the token encoder is built on a general-purpose local LLM at the transmitter rather than a task-specific foundation model.
The newly developed \alg trains a reconfigurable lightweight adapter that enables the LLM to perform token encoding under diverse channel qualities. More comprehensive evaluations of the framework are also provided.

The rest of this paper is organized as follows.
In Sec.~\ref{sec_prelim}, we model the generative semantic communication system in challenging wireless environments.
In Sec.~\ref{sec_sys_framework}, we present the \sys framework and formulate the semantic recovery problem.
In Sec.~\ref{sec_framework}, we propose \alg to solve this problem.
We describe the implementation of \sys in Sec.~\ref{sec_experiment_setup} and provide the simulation results in Sec.~\ref{sec_experiment_results}.
Finally, we conclude the paper and discuss potential extensions in Sec.~\ref{sec_conclusion}.

%% file: sections/4_preliminaries.tex
\section{System Model}
\label{sec_sys_mode}
\label{sec_prelim}

We first model the generative semantic communication system for images and then describe the packet-erasure channel model for challenging wireless environments.

\subsection{Generative Semantic Communication Systems for Images}
\label{ssec_token_communication}

In this paper, we consider a generative semantic communication system over a challenging wireless environment as shown in Fig.~\ref{fig_system}, where the transmitter intends to deliver a set of images to the receiver.
To reduce the bandwidth requirement, the textual description of the semantic meaning of each source image is extracted and transmitted as a token sequence over the wireless channel.
At the receiver, the received tokens are converted back to the textual description, which then drives the text-to-image generative model at the receiver to generate a reconstruction of the source image.

Let $\bm J\in\mathcal I$ denote the target image in image space $\mathcal I$, and let $\bm\alpha\in\mathcal A$ denote its textual description in text space $\mathcal A$.
The mapping between a textual description and its target image is denoted by $\bm g$, i.e., $\bm J=\bm g(\bm\alpha)$, and is captured by a text-to-image generative model in practice.
Following the paradigm of token communication~\cite{Qiao25WCM_Token}, $\bm\alpha$ is further represented by its token sequence, in order to maximize the bandwidth efficiency.
For this purpose, the transmitter is equipped with a tokenizer $\bm q$ associated with \emph{vocabulary} $\cV=\{v_1,\ldots,v_T\}$, where each entry $v_t$ is a textual unit, such as a word, subword, or punctuation fragment.
The vocabulary is indexed by $\cT=\{1,\ldots,T\}$, so that each index $t\in\cT$ refers to entry $v_t\in \cV$.
Therefore, the tokenizer segments a textual description into vocabulary entries and represents them by their index sequence, i.e.,
\beq
\label{eq_source_tokenization}
\bm x=\bm q(\bm \alpha)=[x_1,\ldots,x_N]\in\cT^N,
\eeq
where $N$ denotes the sequence length, and each index $x_n$ is referred to as a \emph{token}.
The ordered vocabulary entries, $(v_{x_1},\ldots,v_{x_N})$, can be reassembled into the textual description by the associated detokenizer, which we denote by $\bm q^{-1}(\cdot)$ for simplicity.
As the mapping between a prompt and its token sequence is reversible, i.e., $\bm\alpha=\bm q^{-1}(\bm x)$, transmitting the token sequence is equivalent to transmitting the prompt.

We assume that the transmitter and the receiver are equipped with a matched pair of tokenizer and detokenizer, i.e., both are associated with the same vocabulary, $\cV$.
This assumption is practical due to the following reason:
As the textual description is ultimately consumed by the generative model at the receiver, which is itself driven by a specific tokenizer, it is natural for the transmitter to serialize its descriptions with the same one.
Even if the transmitter does not host that generative model, maintaining a standalone tokenizer--detokenizer pair only costs very limited storage resources (typically no more than a few megabytes), as the pair consists solely of a vocabulary table and a deterministic segmentation rule.

We further assume that the receiver has the generative model capturing the transmitter's underlying text-to-image mapping, $\bm g(\cdot)$.
In this case, if the receiver obtains the intact token sequence $\bm x$, it recovers the prompt and reconstructs the image by $\bm J=\bm g(\bm q^{-1}(\bm x))$, which is consistent with the target image.\footnote{When this mapping is not captured by the receiver's generative model, knowledge alignment between the transmitter and the receiver becomes essential, which is investigated in our work~\cite{Hu25TIP_Dekag,dekap}.}
This corresponds to the ideal case of a lossless channel, which serves as the performance upper bound of the generative semantic communication system.
In challenging wireless environments, however, the received token sequence generally differs from $\bm x$, and the reconstructed image may consequently deviate significantly from $\bm J$.

\ifSingleColumn
    \renewcommand{\figwidth}{0.65\linewidth}
\else
    \renewcommand{\figwidth}{1\linewidth}
\fi

\begin{figure}[t]
  \centering
  \includegraphics[width=\figwidth]{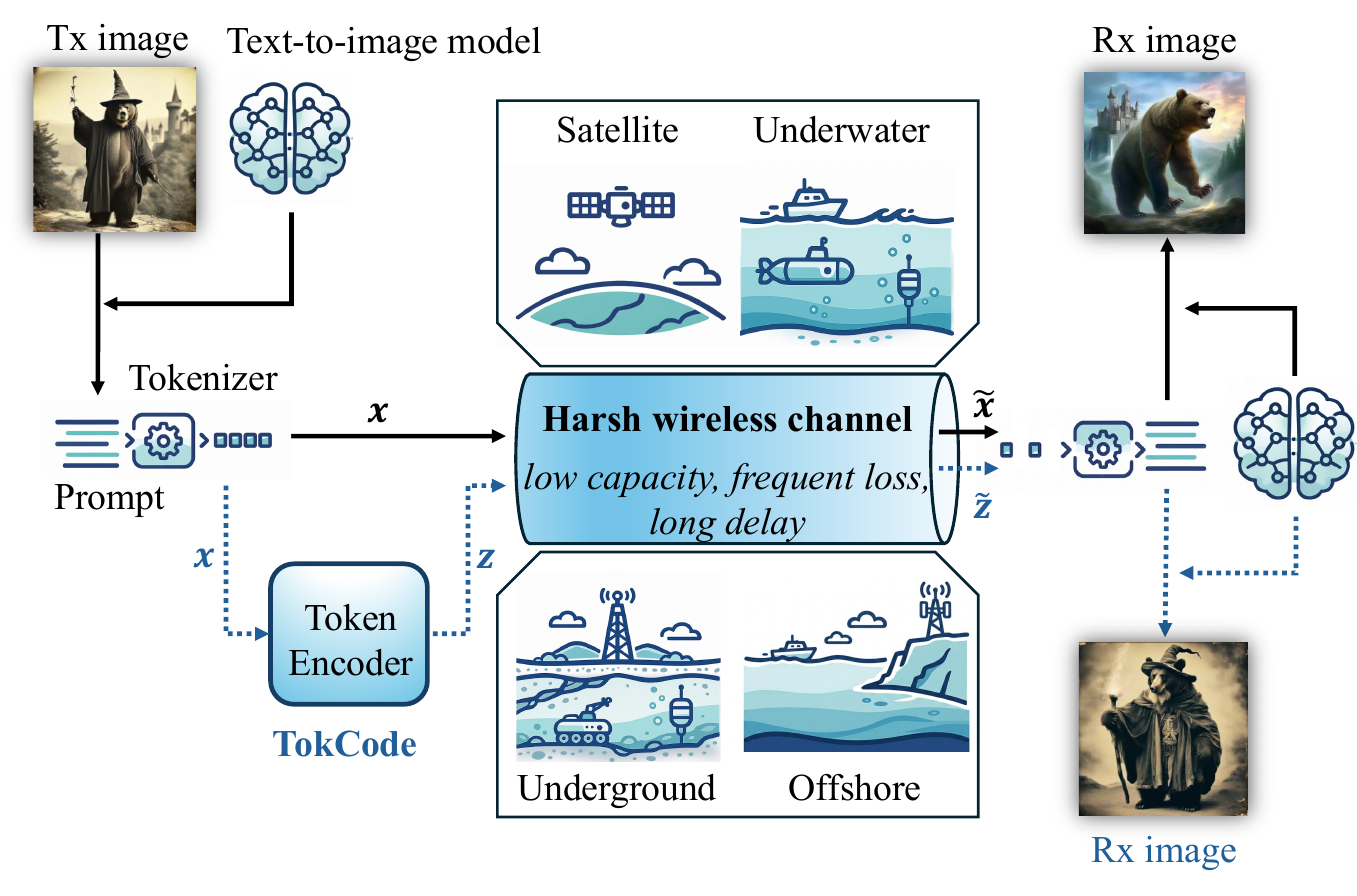}
  \ifSingleColumn
  \else
  \vspace{-2em}
  \fi
  \caption{Token-based generative semantic communication for images over challenging wireless environments.}
  \ifSingleColumn
  \else
  \vspace{-1em}
  \fi
  \label{fig_system}
\end{figure}

\subsection{Channel Model in Challenging Wireless Environments}
\label{ssec_channel_model}

In challenging wireless environments, the channel may suffer from severe fading, shadowing, blockage, interference, and noise bursts, all of which can corrupt data packets.
Meanwhile, given practical latency constraints, long propagation and feedback delays limit the retransmission attempts.
As a result, the receiver suffers from a non-negligible probability of packet loss even with advanced error-control coding.

To facilitate the presentation, we adopt a packet-level channel model that is agnostic to the underlying cause of the impairment.
Since a packet failing the check is directly discarded, the loss occurs at the granularity of a packet~\cite{IEEE802154_2024}: each transmitted packet is either received or erased, and the receiver is aware of which packets are erased.
Such a packet-erasure channel is also commonly adopted in the token-based semantic communication literature~\cite{Qiao25WCM_Token,Han26TMC_Glaris}.
More specifically, we adopt the i.i.d.\ erasure model, in which each packet is erased independently with probability $\gamma\in[0,1]$.
We note that $\gamma$ represents the residual packet-erasure rate after error-control coding and permitted retransmissions.

For the packetization of token sequences, we consider that the transmitter delivers a batch of token sequences together and carries their tokens over a common set of $P$ packets.
The simplest packetization scheme is contiguous packetization, where the token sequences of the batch are allocated into the packets one after another, so that each sequence occupies a single packet or two consecutive ones.
However, such a scheme would expose every textual description to an all-or-nothing loss, where a single packet loss erases the entire descriptions of several images and leaves the receiver with nothing to recover from.
We therefore interleave each token sequence over all the $P$ packets as in~\cite{Cheng24NSDI_GRACE}. 
Specifically, for $\bm x$, its $n$-th token ($n=1,\ldots,N$) is allocated to the $k(n)$-th packet ($k(n) \in \{1,\ldots,P\}$), where
\beq
\label{eq_interleave}
k(n) = 1 + \bigl((n-1)\bmod P\bigr).
\eeq

Denote by $\epsilon_k$ the erasure indicator of the $k$-th packet, which is a random variable following the Bernoulli distribution, $\mathrm{Bernoulli}(\gamma)$.
Then, the channel output corresponding to the input token sequence, $\bm x$, is expressed as
\beq
\label{eq_channel_output}
[\bm h(\bm x|\gamma)]_{n} =
\begin{cases}
x_n, & \text{if } \epsilon_{k(n)}=0,\\
\varnothing, & \text{otherwise},
\end{cases}
\eeq
where $\varnothing$ denotes a missing token, and $\bm h(\cdot|\gamma)$ denotes the packet-erasure channel under erasure rate $\gamma$.
We denote the received token sequence by $\tilde{\bm x}=\bm h(\bm x|\gamma)$ and refer to $\bm\epsilon=(\epsilon_1,\ldots,\epsilon_P)$ as the erasure pattern.

Consequently, the channel condition is characterized by the packet-erasure rate, $\gamma$, and we denote the set of potential packet-erasure rates by $\cR$.
The value of $\gamma$ governs how much of a token sequence survives the wireless channel and reaches the receiver, as well as the relative positions of the surviving tokens and the lengths of the erased spans.
In contrast to common environments, $\gamma$ can be large in challenging wireless environments, e.g., more than half of the tokens of a sequence may be randomly erased on average.

Across $\cR$, the requirement that a token sequence has to meet for robustness varies substantially.
When $\gamma$ approaches one, only one packet survives in most cases, so that the receiver observes a single interleaved sub-sequence of the transmitted tokens.
Robustness then demands that each individual sub-sequence conveys the overall semantics of the textual description on its own.
When $\gamma$ is moderate, several packets may survive and the receiver observes a random combination of their sub-sequences, with the surviving tokens retaining their original relative positions.
Robustness then demands that the sub-sequences are mutually complementary rather than repetitive, so that an arbitrary combination of them stably supplies as much of the original semantics as possible.

However, the above robustness requirements are hardly met by the original token sequence obtained from an ordinary textual description.
Such a sequence inherently carries little redundancy, and any of its tokens may be critical to the overall meaning.
Therefore, the random packet loss can easily lead to semantic distortion beyond recovery at the receiver, which necessitates an encoding process of token sequences before their transmission over challenging wireless channels.

%% file: sections/5_token_encoding_framework.tex
\section{\sys Framework}
\label{sec_sys_framework}

In this section, we introduce a token encoding framework (\sys) for semantic recovery over challenging wireless channels.
We first establish the process of token encoding and then formulate a semantic recovery optimization problem.

\subsection{Token Encoding Process}
\label{ssec_token_encoding_process}

Motivated by the limitations of receiver-side semantic recovery, \sys proactively encodes the source token sequence at the transmitter before channel transmission, as illustrated by the dashed lines in Fig.~\ref{fig_system}.
Instead of sending $\bm x$ directly, the transmitter applies token encoding using the token encoder $\bm f$, i.e.,
\beq
\label{eq_encode}
\bm z=\bm f(\bm x; \bm\Theta)
      =[z_1,\ldots,z_M]\in\cT^M,
\eeq
where $\bm f(\cdot;\bm\Theta)$ denotes the token encoder with parameters $\bm\Theta$,\footnote{For presentation simplicity, we use $\bm\Theta$ as a shorthand for the collection of the token encoder's parameter matrices, which may have diverse dimensions.} and $M$ is the length of the encoded sequence.
Each encoded token still belongs to the same token-index set $\cT$ as the source sequence and corresponds to the same vocabulary~$\cV$.

Moreover, \sys confines the encoded sequence to the token budget of the source sequence, i.e., $M\leq N$, due to the following three reasons.
\emph{First}, it preserves the bandwidth efficiency of token communication.
Since the encoded sequence draws on the same vocabulary $\cV$ and never occupies more token slots than the source sequence, token encoding requires neither an auxiliary codebook nor additional channel resources.
\emph{Second}, it keeps the existing packetization structure intact.
As the encoded sequence never exceeds the $N$ token slots that the source sequence occupies, it can be carried by the same $P$ packets. 
\emph{Third}, it confines the design problem to encoding alone.
Permitting $M>N$ would mix token encoding with the task of adding redundant tokens.
Meanwhile, although $M<N$ is allowed, it is not advantageous since every unused slot reduces the number of tokens that may survive.
The encoder is thus effectively compelled to redistribute the semantic information of the textual description across the $N$ available token slots, which isolates the essential question of how semantics should be spread over a token sequence to survive random erasure.

As analyzed in Sec.~\ref{sec_intro}, realizing the token encoder in~\eqref{eq_encode} demands an accurate understanding of semantics at every level, from individual tokens to their interdependencies and to the sequence as a whole.
\sys exploits the fact that this understanding is precisely what an LLM intrinsically possesses, and instantiates the token encoder $\bm f$ by the local LLM hosted at the transmitter.
The assumption of having an available local LLM is becoming increasingly practical as LLMs continue to shrink in size while end devices grow more capable in computation and storage~\cite{Liu24ICML_MobileLLM,Xu25ASPLOS_OnDeviceLLM}.
Instantiating $\bm f$ in this way, however, entails a tokenizer mismatch: the tokenizer and vocabulary of the local
LLM may differ from those adopted in the generative semantic communication system.
We denote the tokenizer of the local LLM by $\widehat{\bm q}(\cdot)$, its vocabulary by $\widehat{\cV}$, and the corresponding index set by $\widehat{\cT}$.

\sys accommodates this mismatch by allowing the token encoder to operate on the textual description $\bm\alpha$ rather than on $\bm x$ directly.
It takes $\bm x'=\widehat{\bm q}(\bm\alpha)\in(\widehat{\cT})^{N'}$ as its input and produces an encoded sequence $\bm z'\in(\widehat{\cT})^{M'}$, where lengths $N'$ and $M'$ are measured in $\widehat{\cT}$ and generally differ from $N$.
Since $\bm\alpha=\bm q^{-1}(\bm x)$ holds for the source sequence, the token encoder in~\eqref{eq_encode} is realized as the composition
\beq
\label{eq_encoder_composition}
\bm f(\cdot\,;\bm\Theta)
=\bm\varGamma_{N}\circ\bm q\circ\widehat{\bm q}^{\,-1}\circ\bm\phi_{\bm\Theta}\circ\widehat{\bm q}\circ\bm q^{-1},
\eeq
where $\bm\phi_{\bm\Theta}$ denotes the encoding policy of the local LLM, and $\bm\varGamma_{N}(\cdot)$ truncates a sequence to its first $N$ tokens.
Owing to the truncation, the encoded sequence, $\bm z$, always satisfies $M\leq N$, regardless of the intermediate lengths $N'$ and $M'$.

To enable the LLM to perform token encoding, \sys attaches to it a lightweight pluggable adapter, which preserves the model's prior semantic understanding while steering it toward producing erasure-resilient token sequences.
Therefore, leveraging the low-rank adaptation technique~\cite{Hu22ICLR_LoRA} widely used for LLM fine-tuning, we express parameters $\bm\Theta$ in~\eqref{eq_encode} as
\begin{equation}
\label{eq_param_lora}
\bm\Theta = {\bm\Theta}_{\mathrm{LLM}} +  \Delta\bm\Theta= {\bm\Theta}_{\mathrm{LLM}} + \bm B \bm A,
\end{equation}
where $\bm\Theta_{\mathrm{LLM}}\in\mathbb R^{D_{\mathrm{out}}\times D_{\mathrm{in}}}$ denotes a pretrained parameter matrix of the local LLM, with $D_{\mathrm{in}}$ and $D_{\mathrm{out}}$ being the input and output dimensions, respectively,
$\bm B\in\mathbb R^{D_{\mathrm{out}}\times r}$ and $\bm A\in\mathbb R^{r\times D_{\mathrm{in}}}$ are the parameter matrices of the adapter, whose product is of rank at most $r$, and $r$ is selected to be far smaller than $D_{\mathrm{in}}$ and $D_{\mathrm{out}}$.
By freezing ${\bm\Theta}_{\mathrm{LLM}}$ and training the low-rank pair $(\bm B,\bm A)$ alone, the semantic prior of the LLM is retained while the low-rank update steers the LLM toward the token encoding task.

Owing to small rank $r$, the adapter contains substantially fewer parameters than ${\bm\Theta}_{\mathrm{LLM}}$ and thus can be stored at a low cost.
When token encoding is to be performed, the low-rank adapter can be merged into the LLM parameters on demand and subtracted afterwards to restore the LLM to its original state for its original applications.
Such flexibility renders \sys a \emph{pluggable} module of the transmitter.

\subsection{Semantic Recovery Problem}
\label{ssec_objective_definition}

With the token encoding process established, we now formulate the problem that \sys aims to solve through token encoding, i.e., the semantic recovery of the image generated at the receiver under random packet erasure.
Since the transmitted token sequence is replaced by the encoded one, the received sequence becomes $\bm h(\bm z|\gamma)$, and the generated image at the receiver can be expressed as
\begin{equation}
\label{equ_generated_image}
\tilde{\bJ}=\bm g\circ\bm q^{-1}\circ\bm h(\bm z|\gamma),
\end{equation}
where $\circ$ denotes function composition.

Intuitively, the objective of token encoding would be to optimize the adapter so that $\tilde{\bJ}=\bJ$.
However, random token erasure caused by the channel can induce non-negligible semantic distortion in the received token sequences, resulting in substantial structural and visual discrepancies between $\tilde{\bJ}$ and $\bJ$.
Consequently, exact pixel-wise equivalence between $\tilde{\bJ}$ and $\bJ$ is infeasible in practice and is therefore unsuitable as the objective of \sys.
Instead, we follow the multi-metric image-consistency principle in~\cite{Hu25TIP_Dekag}
and quantify the similarity between $\tilde{\bJ}$ and $\bJ$ at two complementary levels, namely visual similarity and semantic similarity.

Specifically, to measure the visual similarity, we employ the DINO model~\cite{Simeoni26TMLR_DINOv3}, which provides high-quality dense visual features, and define $S_{\mathrm{DINO}}(\tilde{\bJ}, \bJ)$ as the cosine similarity between the DINO features of $\tilde{\bJ}$ and $\bJ$.
To measure the semantic similarity, we employ the CLIP model~\cite{Radford21ICML_CLIP}, which embeds images into a semantic space shared with textual descriptions and thereby captures high-level meaning, and define $S_{\mathrm{CLIP}}(\tilde{\bJ}, \bJ)$ as the cosine similarity between the CLIP features of $\tilde{\bJ}$ and $\bJ$.
We refer to the visual and semantic similarities as the \emph{DINO-score} and \emph{CLIP-score}, respectively, and formulate the semantic recovery problem as
\beq
\label{eq_obj_token_enc}
\max_{(\bm B,\bm A)}\;\mathbb E\!\left[S_{\mathrm{DINO}}(\bJ,\tilde{\bJ})+S_{\mathrm{CLIP}}(\bJ,\tilde{\bJ})\right],
\eeq
where the expectation is taken over channel condition $\gamma\in\cR$, random erasure pattern $\bm\epsilon$, and potential target image $\bJ$.

Solving~\eqref{eq_obj_token_enc} is highly challenging mainly because the training of the adapter lacks reliable supervision.
Supervised fine-tuning would require target encoded sequences as labels, which neither exist in available datasets nor can be provided manually, as no explicit rule prescribes what makes a token sequence resilient to random erasure, especially across diverse non-trivial erasure rates.

Although reinforcement learning may provide a viable alternative to train the adapter directly with the reward in~\eqref{eq_obj_token_enc}, it is impeded by the prohibitive cost of the reward evaluation multiplied by the vast number of evaluations that the search demands.
Specifically, a single evaluation of the reward traverses a cascade of deep models, i.e., the autoregressive generation of the LLM-based token encoder, the iterative denoising of the text-to-image model, and the DINO and CLIP feature extraction, which incurs a high computational cost.
Meanwhile, reinforcement learning is expected to confront an obvious plateau strategy, namely repeating the salient tokens, which is easy to find but far from optimal.
A better encoding strategy requires token substitutions coordinated over multiple positions, whereas its intermediate attempts often degrade the reward.
Consequently, a substantial number of trials is necessary before such a strategy can be revealed, which is further inflated by the randomness of the erasure pattern.
Given the high cost of every single evaluation, optimizing the adapter by reinforcement learning alone is hardly affordable.

Furthermore, the diversity of channel conditions aggravates the difficulty of solving~\eqref{eq_obj_token_enc}.
Optimizing a single shared adapter $(\bm B, \bm A)$ for all the channel conditions usually stagnates or converges to a solution biased toward a subset of $\cR$ because the encoding strategies favored under different erasure rates may differ markedly.
In comparison, training a dedicated adapter for every $\gamma\in\cR$ scales both the training and the storage resource consumption linearly with $|\cR|$.
Moreover, if each change of the channel condition required subtracting the integrated low-rank adapter and merging a new one across all the adapted layers, frequent switching would incur considerable computation overhead.

%% file: sections/6_cadet.tex
\section{\alg: Channel-Quality-Aware Distillation for Token Encoder Training}
\label{sec_tokcode_framework}
\label{sec_framework}

We propose a channel-quality-aware distillation approach for token encoder training~(\alg), which expedites the training of the LLM adapter and addresses the challenges of solving~\eqref{eq_obj_token_enc}.
The key idea is to first solve a surrogate problem of~\eqref{eq_obj_token_enc} on a foundation model whose architecture and pretraining make it inherently more suited to token encoding than the LLM at the transmitter.
With this model as an expert, it provides the supervision labels from which the adapter, acting as a student, distills the token-encoding capability.

Specifically, \alg consists of three stages, namely {expert formation}, {expert distillation}, and {student reinforcement}.

%% file: sections/6a_stage1_expert_tuning.tex
\subsection{Expert Formation}
\label{ssec_stage1_expert_tuning}

The first stage of \alg\ forms an expert whose output will supervise the training of the actual token encoder in \sys.
To obtain the expert, \alg\ first selects a ``talent'' model as the basis, and then trains this talent into an expert for token encoding.
There are two requirements for the selection of the talent:
(i) The talent has to share the tokenizer and vocabulary with the communication system.
Token encoding amounts to deciding which token in the vocabulary occupies each position of the transmitted sequence.
Thus, the tokens that the talent selects have to coincide with those actually transmitted, which holds only if the talent operates in the same vocabulary. 
(ii) The talent also has to possess the semantic understanding spanning individual tokens, token interdependencies, and the sequence as a whole, as identified in Sec.~\ref{ssec_token_encoding_process}.
In other words, it has to decide the token of every position with the entire source sequence in view.

Guided by these two requirements, we adopt as the talent an encoder--decoder foundation model pretrained to reconstruct contiguous spans of the text that are masked out from its input.
Such a pretraining task compels the model to infer the missing content from the surrounding context, thereby endowing it with exactly the levels of understanding required above.
Meanwhile, its encoder module exposes the global semantics of the source sequence to every output position, so that each encoded token is produced with the entire source sequence in view.
A decoder-only model, in contrast, is inherently constrained by causal attention, under which every encoded token takes into account its left context alone.

Even with such a well-suited basis, however, training the talent under the objective in~\eqref{eq_obj_token_enc} faces two intertwined difficulties.
\emph{First}, optimizing the image-level objective directly would require propagating gradients across the entire generation process of the image generative model, whose computation graph is prohibitively large.
\emph{Second}, the talent emits a discrete token index at each output position, which severs the gradient propagation path from the objective back to its parameters.

We address the first difficulty by recasting the training objective from the image level to the sentence level.
The intuition is that the closer the received token sequence is to the source one in sentence-level semantics, the more similar the images generated from them are.
To this end, we adopt a \emph{sentence embedder} that maps a sentence into an embedding space in which the angle between embeddings reflects sentence-level similarity.
This sentence embedder is required to be associated with the same tokenizer $\bm q(\cdot)$ and vocabulary $\cV$, so that the output of the talent can be fed to it directly, without an intermediate conversion into text that would block gradient back-propagation.
Meanwhile, evaluating the semantic similarity at the sentence level keeps the image generative model out of the computation graph altogether.

Accordingly, we propose the following sentence-level surrogate objective for~\eqref{eq_obj_token_enc}:
\beq
\label{eq_obj_sentence_total}
\cL_{\mathrm{sent}}
=
\mathbb E
\Bigl[
-S_{\mathrm{sent}}(\bm x, \tilde{\bm z})
+\lambda\, D_{\mathrm{norm}}(\bm x, \tilde{\bm z})
\Bigr],
\eeq
where the expectation is taken over the potential source images and the random erasure patterns.
Here, $\bm x$ is the token sequence of the textual description of target image $\bm J$, and $\tilde{\bm z}$ is the encoded sequence, $\bm z$, produced by the talent after passing the erasure channel.
Both sequences are expressed in the vocabulary, $\cV$, of the system.
In addition, $S_{\mathrm{sent}}$ is the cosine similarity between the source and received sentence embeddings extracted by the sentence embedder, and $D_{\mathrm{norm}}$, weighted by $\lambda\geq 0$, is the squared difference between the $\ell_2$ norms of the two embeddings capturing the magnitude discrepancy.
Denote the sentence embedder by $\bm\xi(\cdot)$, and the above two terms are expressed as
\beq
\label{eq_obj_sentence_sent}
\begin{aligned}
S_{\mathrm{sent}}(\bm x, \tilde{\bm z})
&=
\cos\!\left(
\bm\xi(\bm x),\bm\xi(\tilde{\bm z})
\right),\\
D_{\mathrm{norm}}(\bm x, \tilde{\bm z})
&=
\left(
\|\bm\xi(\tilde{\bm z})\|_2-\|\bm\xi(\bm x)\|_2
\right)^2.
\end{aligned}
\eeq

We then address the second difficulty by exploiting the vocabulary shared by the talent and the embedder.
Since both index the same vocabulary, $\cV$, we construct an embedding-space shortcut that bypasses the discretization step.
Specifically, at each output position $m$, the talent produces a logit vector $\bm l_m\in\bR^T$, and the probability vector to select among all the $T$ tokens is given by
\beq
\label{eq_soft_prob}
\bm r_m=\mathrm{softmax}(\bm l_m)\in\bR^{T}.
\eeq

Normally, the token output at position $m$ is taken as the index of the largest element of $\bm r_m$, and the embedder then uses this hard index to retrieve the corresponding row from its token embedding table $\bm V_{\bm\xi}\in\bR^{T\times D'}$.
Beyond this hard selection, $\bm r_m$ can also serve as a soft index that combines all $T$ rows of $\bm V_{\bm\xi}$.
We refer to the two resulting vectors as the hard and soft token embeddings, i.e.,
\beq
\label{equ_soft_hard}
\bm v_m^{\mathrm{hard}}=\bm V_{\bm\xi}[z_m,:],\quad
\bm v_m^{\mathrm{soft}}=\bm r_m^{\top}\bm V_{\bm\xi},
\eeq
where $z_m=\arg\max_{j\in\cT}\, r_{m,j}$ is the hard token index.

The soft embedding alone, however, cannot be fed to the sentence embedder.
Being a combination of all rows of $\bm V_{\bm\xi}$, it generally coincides with none of them and is therefore out of the input distribution on which the sentence embedder is trained.
The sentence embedding extracted from such an input would be unreliable, making the resulting similarity a misleading signal for training the talent.
To retain the genuine token embedding in the forward computation while still exposing a differentiable path, we apply the straight-through estimator (STE)~\cite{STE}, which combines the two embeddings as
\beq
\label{eq_ste_emb}
\bm v_m^{\mathrm{ste}}
=
\mathrm{sg}\!\left(
\bm v_m^{\mathrm{hard}}-\bm v_m^{\mathrm{soft}}
\right)
+\bm v_m^{\mathrm{soft}},
\eeq
where the stop-gradient operator $\mathrm{sg}(\cdot)$ acts as the identity in the forward pass and returns zero gradient in the backward pass.
Consequently, the forward pass yields $\bm v_m^{\mathrm{ste}}=\bm v_m^{\mathrm{hard}}$, and the backward pass sees only $\bm v_m^{\mathrm{soft}}$, whose gradient propagates through $\bm r_m$ to the trainable parameters of the talent.

To further improve the efficiency of training, we adopt three special designs.
We first reduce the number of steps that the gradient has to traverse.
Rather than adopting the default autoregressive decoding, we let the talent produce the whole encoded sequence $\bm z$ within a single forward pass.
In this way, we avoid unrolling a computation graph of sequentially encoding $N$ tokens.
Then, we reduce the dimension of the parameter space to be searched by adapting the talent with a low-rank adaptation analogous to~\eqref{eq_param_lora}.
Furthermore, we train the talent under a curriculum, which first fits it to reproduce its own input as a warm-up and optimizes $\cL_{\mathrm{sent}}$ in~\eqref{eq_obj_sentence_total} afterwards.
Since the identity encoding is an optimal solution under a lossless channel, it provides a far stronger initialization for the subsequent optimization than the pretrained talent.

A final aspect to address is that different erasure rates generally favor markedly different token encoding strategies.
Training a single expert to cover the whole range $\cR$ would encounter conflicting objectives, which often lead to stalled progress or biased convergence.
Fortunately, an expert is only used to generate the supervision labels.
Therefore, we can train multiple experts, each under a different channel condition, without incurring additional storage overhead in the deployment of \sys.
In particular, we select three representative erasure rates from $\cR$, namely a high one $\gamma_{\mathrm{H}}$, a moderate one $\gamma_{\mathrm{M}}$, and a low one $\gamma_{\mathrm{L}}$, and train a dedicated expert under each of them.
The three experts formed in this stage are denoted by $\bm\zeta_{\mathrm{H}}(\cdot)$, $\bm\zeta_{\mathrm{M}}(\cdot)$, and $\bm\zeta_{\mathrm{L}}(\cdot)$, respectively.


%% file: sections/6b_stage2_distillation.tex
\subsection{Expert Distillation}
\label{ssec_stage2_distillation}

The second stage of \alg\ transfers the token-encoding capability of the obtained experts into the adapter to be deployed at the transmitter's local LLM.
Following the standard distillation terminology, we refer to the experts as the \emph{teachers} and to the adapted LLM at the transmitter as the \emph{student}.
The key advantage of this transfer lies in converting the challenging token sequence optimization underlying~\eqref{eq_obj_token_enc} into supervised learning.
The experts formed in Sec.~\ref{ssec_stage1_expert_tuning} supply exactly the labels needed for the supervision:
their encoded sequences serve as token-level targets, with which the student can be trained by next-token prediction.

The first step to realize this transfer is constructing the datasets for distillation.
To this end, each of $\bm\zeta_{\mathrm{H}}(\cdot)$, $\bm\zeta_{\mathrm{M}}(\cdot)$, and $\bm\zeta_{\mathrm{L}}(\cdot)$ encodes every textual description in $\cD_{\mathrm{pmt}}$.
Here, $\cD_{\mathrm{pmt}}$ is a text dataset specifically prepared for the distillation, comprising diverse text-to-image prompts of varying lengths.
The label is the output token sequence produced by the expert.
However, the expert produces its encoded sequence in the token domain specified by $\bm q(\cdot)$ and $\cV$, whereas the student generates in its own domain associated with $\widehat{\bm q}(\cdot)$ and $\widehat{\cV}$.
To resolve this mismatch, we use text as a bridge: each encoded sequence of an expert is detokenized into text and stored as the supervision label, which the student re-tokenizes with $\widehat{\bm q}(\cdot)$ to obtain the target sequence.
Running the three experts over the same prompt dataset accordingly yields three datasets, i.e.,
\beq
\label{eq_distill_dataset}
\begin{aligned}
\cD_{\mathrm{X}}
\!=\!\bigl\{({\bm x}', {\bm y}')\,\big|\;&
   {\bm x}' = \widehat{\bm q}(\bm\alpha),\\[-1mm]
   &{\bm y}' = \widehat{\bm q}\circ\bm q^{-1}\!\circ\bm\zeta_{\mathrm{X}}(\bm\alpha),\quad
   \forall \bm\alpha\!\in\!\cD_{\mathrm{pmt}}\bigr\},
\end{aligned}
\eeq
where subscript $\mathrm{X}$ stands for any of $\mathrm{L}$, $\mathrm{M}$, and $\mathrm{H}$, and both the input, ${\bm x}'$, and the label, ${\bm y}'$, are in $\widehat{\cT}$, i.e., the token domain in which the student is trained.

The three datasets defined in~\eqref{eq_distill_dataset} differ only in their labels, each set of which represents the token-encoding strategy of one expert.
However, unlike the experts, which stay offline, the adapter is deployed at the transmitter, where a dedicated adapter per erasure rate would scale up the local storage and incur a large switching overhead whenever the channel condition varies.
To accommodate the three strategies within one adapter while mitigating their conflict, we adopt the lightweight rank-reweighting mechanism in~\cite{Hu25TIP_Dekag} to construct a reconfigurable low-rank adapter.

Specifically, instead of training a dedicated adapter under every erasure rate, we keep a single pair $(\bm B,\bm A)$ shared by all the erasure rates and let $\gamma$ reconfigure the scales of its rank-one components, which yields
\beq
\label{eq_gated_lora_update}
\Delta \bm\Theta
=\bm B\,\mathrm{Diag}\!\left(\bm\beta(\gamma)\right)\bm A
=\sum_{i=1}^{r}\beta_{i}(\gamma)\,\bm b_{i}\bm a_{i}^{\mathsf{T}},
\eeq
where $\mathrm{Diag}(\cdot)$ denotes the diagonal matrix formed from its argument, $\bm b_{i}$ and $\bm a_{i}^{\mathsf{T}}$ denote the $i$-th column of $\bm B$ and the $i$-th row of $\bm A$, respectively, and $\bm\beta:\cR\to\bR^{r}$ is a lightweight neural model, referred to as the \emph{reweighting function}, which maps the erasure rate to the scaling vector $[\beta_{1}(\gamma),\ldots,\beta_{r}(\gamma)]$.

The motivation behind~\eqref{eq_gated_lora_update} can be explained as follows.
The low-rank adapter is a sum of $r$ rank-one components, each of which can be interpreted as carrying one piece of token-encoding knowledge, e.g., repeating a salient token at a proper position or preserving the descriptive details of the source prompt.
By reweighting them, the same group of components can potentially serve the demands of all the erasure rates.
The conflict among the strategies is thus handled explicitly by the reweighting factors in $\bm\beta(\gamma)$.

More specifically, the reweighting function consists of three layers.
The first layer maps the scalar $\gamma$ into a sinusoidal embedding of multiple frequencies using the construction introduced for positional encoding in~\cite{Vaswani17NeurIPS_Attention}.
The following two layers are dense linear layers, which together map the embedding to the $r$ scaling factors.
Each adapter is equipped with its own reweighting function, whose number of parameters remains negligible compared with that of the low-rank matrices due to its simple structure.
The switching overhead is thereby avoided as well since a varying channel condition only requires updating the $r$ scaling factors.

With the labels in~\eqref{eq_distill_dataset} and the parameterization in~\eqref{eq_gated_lora_update} established, the second stage of \alg\ is formulated as the following channel-quality-aware distillation problem:
\beq
\label{eq_distill_nll}
\min_{\bm B,\bm A,\,\bm\beta}
\sum_{\mathrm{X}\in\{\mathrm{H,M,L}\}}
\sum_{({\bm x}',{\bm y}')\in\cD_{\mathrm{X}} }
\sum_{n=1}^{|{\bm y}'|}
\ell_{\mathrm{CE}}\!\left(
{y}'_{n},\,
\left[\bm \phi_{\bm\Theta_{{\mathrm{X}}}}\!\left({\bm x}', {\bm y}'_{<n}\right)\right]_n
\right)\!,
\eeq
where $\bm\phi_{\bm\Theta}$ is the encoding policy of the local LLM defined in~\eqref{eq_encoder_composition}, $\ell_{\mathrm{CE}}(\cdot)$ denotes the cross-entropy loss, ${y}'_{n}$ is the $n$-th token of expert-generated label ${\bm y}'$, ${\bm y}'_{<n}$ collects those preceding it, $[\cdot]_n$ denotes the predicted distribution of the student at the $n$-th position, and $\bm\Theta_{{\mathrm{X}}}=\bm\Theta_{\mathrm{LLM}}+ \Delta \bm\Theta|_{\gamma = \gamma_{\mathrm X}}$.
Problem~\eqref{eq_distill_nll} is a next-token prediction problem.
As the target tokens are known in advance, the cross-entropy losses within a sequence can be computed in parallel, and their gradients are free of the variance caused by a long autoregressive rollout.

%% file: sections/6c_stage3_rl_refinement.tex
\subsection{Student Reinforcement}
\label{sec_student_rl_refinement}

In the final stage, the student LLM is refined by reinforcement learning with the real objective function in~\eqref{eq_obj_token_enc} as the reward, so that its token encoding is further aligned with the image-level objective.
Following the standard reinforcement learning terminology, we regard the source sequence $\bm x'$ together with the erasure rate $\gamma$ as the \emph{state}, and the sequence $\bm z'$ generated by the student as the \emph{action} taken at that state.
Under this view, the student, i.e., $\bm\phi_{\bm\Theta}$, can be viewed as a \emph{policy} that maps a state to a distribution over actions.
When an action is taken at a state, it leads to a scalar reward, and reinforcement learning will improve the policy to maximize the expected reward.
To align the student refinement with the image-level objective in~\eqref{eq_obj_token_enc}, it is natural to take the evaluated value of the objective function as the reward.
We note that such a reward is highly stochastic.
An action is first converted into the encoded sequence by $\bm z = \bm\varGamma_{N}\circ\bm q\circ\widehat{\bm q}^{\,-1}(\bm z')$ according to~\eqref{eq_encoder_composition} and then undergoes the random packet-erasure channel in~\eqref{eq_channel_output}, which yields $\tilde{\bm z}=\bm h(\bm z|\gamma)$.
It is $\tilde{\bm z}$, instead of the action itself, that determines the generated image and thereby the value of the objective function in~\eqref{eq_obj_token_enc}.

We solve this reinforcement learning problem with group-relative policy optimization (GRPO) \cite{Shao24arXiv_GRPO}, which is tailored to reinforcement learning of LLMs on complex tasks~\cite{Guo25arXiv_DeepSeekR1}. 
Unlike conventional methods that learn a separate value model, GRPO directly compares a group of actions sampled by the policy at the same state to determine the direction of policy improvement, thereby reducing cost and improving stability.

To apply GRPO to the student reinforcement of \alg, we introduce the following notations.
Since the student reuses a local LLM that generates autoregressively, $\bm\phi_{\bm\Theta}$ can be equivalently factorized into a sequence of next-token policies. 
Specifically, for the $n$-th position, the next-token policy is expressed as $\bm \phi_{\bm\Theta}(\bm x',\bm z'_{<n})\in\bR^{\widehat{T}}$, which takes the source sequence $\bm x'$ and the previously generated tokens $\bm z'_{<n}=[z'_1,\ldots,z'_{n-1}]$ as context and outputs a probability vector over $\widehat{T}$.
We note that the erasure rate, $\gamma$, enters through the reconfigurable low-rank parameterization in~\eqref{eq_gated_lora_update}, so that $\bm\Theta$ is itself a function of $\gamma$. 
At inference, the $n$-th encoded token is taken as the index of the maximum element in $\bm \phi_{\bm\Theta}(\bm x',\bm z'_{<n})$, whereas during reinforcement learning the token is randomly sampled according to the probabilities in order to enable exploration.

Given a sampled state $(\bm x',\gamma)$, we first draw a group of $G$ actions $\{\bm z^{\prime(g)}\}_{g=1}^{G}$ by token-wise sampling.
Each of them is converted into an encoded sequence associated with $\cV$ by $\bm z^{(g)} = \bm\varGamma_{N}\circ\bm q\circ\widehat{\bm q}^{\,-1}(\bm z^{\prime(g)})$.
Under erasure rate $\gamma$, all $G$ encoded sequences are then passed through the same packet-erasure realization, i.e., a single fixed erasure pattern shared across the group, and the reward is evaluated for each, obtaining $R^{(1)},\ldots,R^{(G)}$.
Here, sharing one erasure realization within the group ensures that the reward differences among the sampled sequences stem solely from the encoding policy.

For each sampled sequence, GRPO then computes an advantage value that measures the excess of its reward over the group average. 
We denote the advantage of the $g$-th sampled action, i.e., $\bm z^{\prime(g)}$, by $\Lambda^{(g)}$, which is obtained by standardizing its reward within the group, i.e.,
\beq
\label{eq_student_advantage}
\Lambda^{(g)}
=({R^{(g)}-\mu_{R}})/{\sigma_{R}},
\eeq
where $\mu_{R}$ and $\sigma_{R}$ denote the mean and the standard deviation of the $G$ rewards within the group, respectively.

In GRPO, $\bm\Theta$ is updated towards raising the probability of the sampled sequences with positive advantages in~\eqref{eq_student_advantage} and lowering that of the ones with negative advantages while limiting the magnitude of the policy change to avoid instability.
Rather than operating on the actual probability of an encoded sequence, which is prone to vanish due to the multiplication of $N$ token probabilities, GRPO operates on the generation probability of each token.
As the predictability of different tokens may still differ substantially, GRPO normalizes the generation probability of every sampled token by its value under the old policy that has produced the sample.
In particular, for the $n$-th token of the $g$-th sequence, the resulting ratio is expressed as
\beq
\rho_n^{(g)}
=
\frac{\bigl[\bm \phi_{\bm\Theta}\!\bigl(\bm x',\bm z^{\prime(g)}_{<n}\bigr)\bigr]_{z^{\prime(g)}_n}}
{\bigl[\bm \phi_{\bm\Theta_{\mathrm{old}}}\!\bigl(\bm x',\bm z^{\prime(g)}_{<n}\bigr)\bigr]_{z^{\prime(g)}_n}},
\label{eq_student_ratio}
\eeq
where $[\cdot]_{z'}$ takes the element of a probability vector corresponding to token $z'\in\widehat{\cT}$,
and $\bm \phi_{\bm\Theta_{\mathrm{old}}}$ indicates the policy that produced the samples, which are frozen while $\bm\Theta$ is updated. 

Maximizing $\rho_n^{(g)}\Lambda^{(g)}$ hence raises the probability of emitting $z^{\prime(g)}_n$ if $\Lambda^{(g)}>0$ and lowers it if $\Lambda^{(g)}<0$.
Meanwhile, $\rho_n^{(g)}$ reflects the policy shift and should be limited for stability.
Accordingly, the utility to be maximized is defined as
\beq
\mathcal U
=
\sum_{g=1}^{G}\sum_{n=1}^{N}
{\min\Bigl(
\rho_n^{(g)}\Lambda^{(g)},\,
\mathrm{clip}\bigl(\rho_n^{(g)},1-\varepsilon,1+\varepsilon\bigr)\cdot \Lambda^{(g)}
\Bigr)\over GN},
\label{eq_student_utility}
\eeq
where $\mathrm{clip}(\cdot,1-\varepsilon,1+\varepsilon)$ restricts the ratio to region $[1-\varepsilon,1+\varepsilon]$ with clipping threshold $\varepsilon\in(0,1)$.

Finally, to prevent the exploration in reinforcement learning from substantially degrading the encoding capability established in Stage~2, we anchor the current policy to the reference policy $\bm\phi_{\bm\Theta_{\mathrm{ref}}}$, which is the outcome of the expert distillation stage.
This is realized by penalizing $\mathcal U$ with the Kullback--Leibler (KL) divergence of the current policy from $\bm\phi_{\bm\Theta_{\mathrm{ref}}}$. 
Consequently, at each step of reinforcement training of the student, the policy is updated by solving
\beq
\max_{\bm B,\bm A,\,\bm\beta}~
\mathcal U-\eta\,D_{\mathrm{KL}}\!\bigl(\bm\phi_{\bm\Theta}\,\|\,\bm\phi_{\bm\Theta_{\mathrm{ref}}}\bigr),
\label{eq_student_total_obj}
\eeq
where $\eta>0$ weights between the maximization of utility $\mathcal U$ and the anchoring.
For each group of samples, the parameters are updated by a few steps, after which $\bm\Theta_{\mathrm{old}}$ is resynchronized to the newest parameters and the next group is drawn.
At each step, a random textual description $\bm\alpha$ is drawn from the prompt dataset $\cD_{\mathrm{pmt}}$, and the erasure rate is drawn from the three representative ones, i.e., $\gamma\in\{\gamma_{\mathrm{L}},\gamma_{\mathrm{M}},\gamma_{\mathrm{H}}\}$.

Restricting $\gamma$ to the three representative rates, instead of the whole range, $\cR$, is essential to the stability of the reinforcement.
In~\eqref{eq_student_advantage}, the reward is evaluated on the received sequence corrupted by the random channel, whose effect is averaged out only when the same erasure rate is frequently visited.
If $\gamma$ were drawn from $\cR$, the visits would scatter over a continuum of rates and hardly any would be revisited.

Such a design does not prevent the trained adapter from serving an arbitrary $\gamma\in\cR$.
After the student reinforcement, the adapter reweighted by $\gamma_{\mathrm{H}}$, $\gamma_{\mathrm{M}}$, and $\gamma_{\mathrm{L}}$ is evaluated offline to obtain $V(\gamma_{\mathrm{X}},K)$, which represents the average image-level similarity in~\eqref{eq_obj_token_enc} achieved by the encoded sequence when $K\leq P$ of the $P$ packets randomly survive.
Meanwhile, under the erasure model in Sec.~\ref{ssec_channel_model}, the impact of $\gamma$ is characterized by the distribution of the number of surviving packets, i.e., $\Pr\{K\,|\,\gamma\}=\binom{P}{K}(1-\gamma)^{K}\gamma^{P-K}$.
Therefore, the reweighting input, denoted by $p$, can be selected by
\beq
\label{eq_gate_selection}
p^*(\gamma)
=\arg\max_{\gamma_\mathrm{X}\in\{{\gamma_\mathrm{L}},{\gamma_\mathrm{M}},{\gamma_\mathrm{H}}\}}
\sum_{K=1}^{P}\Pr\{K\,|\,\gamma\}\cdot V(\gamma_{\mathrm{X}},K).
\eeq

This design keeps the performance at every $\gamma\in\cR$ predictable, rather than fully relying on the adapter to generalize to the erasure rates unseen in the training.

The complete \alg\ is summarized as Algorithm~\ref{alg_cmoe_grpo}.

\input{sections/alg_2.tex}

%% file: sections/alg_2.tex
\begin{algorithm}[t]
\ifSingleColumn
\footnotesize
\else
\small
\fi
\caption{\alg\ for training the token encoder.}
\label{alg_cmoe_grpo}
\begin{algorithmic}[1]
\Require Prompt dataset $\cD_{\text{pmt}}$; representative erasure rates $\cR_{\mathrm{r}}=\{\gamma_{\mathrm{L}},\gamma_{\mathrm{M}},\gamma_{\mathrm{H}}\}$; pretrained talent; transmitter-side local LLM; pretrained sentence embedder $\bm\xi(\cdot)$; text-to-image generative model; surrogate loss weight $\lambda$; group size $G$.
\Ensure Optimized adapter for LLM, including low-rank parameters $(\bm B^*,\bm A^*)$ and reweighting function $\bm\beta^*$.
\vspace{1ex}
\For{each rate $\gamma_{\mathrm{X}}\in\cR_{\mathrm{r}}$}
  \State Warm up a low-rank adapter of talent on identity encoding.
  \State Minimize $\cL_{\mathrm{sent}}$ in~\eqref{eq_obj_sentence_total} through the STE path in~\eqref{eq_ste_emb}, and obtain expert $\bm\zeta_{\mathrm{X}}(\cdot)$.
\EndFor
\For{each rate $\gamma_{\mathrm{X}}\in\cR_{\mathrm{r}}$}
  \State Encode $\cD_{\text{pmt}}$ by $\bm\zeta_{\mathrm{X}}(\cdot)$ and detokenize the outputs into the textual labels, forming $\cD_{\mathrm{X}}$ in~\eqref{eq_distill_dataset}.
\EndFor
\For{each distillation step}
  \State Draw $\mathrm{X}$ uniformly and a training batch from $\cD_{\mathrm{X}}$, and update $(\bm B,\bm A,\bm\beta)$ in~\eqref{eq_gated_lora_update} by minimizing~\eqref{eq_distill_nll}.
\EndFor
\State Freeze the distilled student parameters as reference policy $\bm\phi_{\bm\Theta_{\mathrm{ref}}}$.
\For{each reinforcement step}
  \State Draw $\bm \alpha \in\cD_{\text{pmt}}$ and $\gamma_{\mathrm{X}}\in\cR_{\mathrm{r}}$, and sample $G$ actions $\{\bm z^{\prime(g)}\}$ from $\bm\phi_{\bm\Theta_{\mathrm{old}}}$.
  \State Pass them through erasure channel with a shared realization across the group to obtain $\{\tilde{\bm z}^{(g)}\}$.
  \State Render the received texts into images, and evaluate rewards $\{R^{(g)}\}$ and advantages $\{\Lambda^{(g)}\}$ in~\eqref{eq_student_advantage}.
  \State Optimize~\eqref{eq_student_total_obj} w.r.t. $\Delta\bm\Theta = \bm B\,\mathrm{Diag}\!\left(\bm\beta(\gamma_{\mathrm{X}})\right)\bm A$ for a few steps, and set $\bm\Theta_{\mathrm{old}}\!\leftarrow\!\bm\Theta_{\mathrm{LLM}}+\Delta\bm\Theta$.
\EndFor
\State Evaluate $V(\gamma_{\mathrm{X}},K)$ for $\gamma_{\mathrm{X}}\in\cR_{\mathrm{r}}$ and tabulate $p^{\star}(\gamma)$ by~\eqref{eq_gate_selection}.
\end{algorithmic}
\end{algorithm}

%% file: sections/7a_experiments_setup.tex
\section{System Implementation}
\label{sec_experiment_setup}

Below, we first specify the simulation scenario together with the used dataset, and then detail how the token encoder of \sys is instantiated and trained by \alg.

\subsection{Simulation Scenario and Dataset}
\label{ssec_scenario_dataset}

We instantiate the generative semantic communication system of Sec.~\ref{ssec_token_communication} for the image transmission task between a transmitter and a receiver.
Their text-to-image mapping, $\bm g(\cdot)$, is realized by PixArt-Sigma~\cite{Chen24ECCV_PixArtSigma}, a diffusion transformer that generates an image from the textual condition extracted by its T5-family text encoder~\cite{Raffel20JMLR_T5}.
Accordingly, the tokenizer, $\bm q(\cdot)$, detokenizer, $\bm q^{-1}(\cdot)$, and vocabulary, $\cV$, of the system are also those of T5-XXL, as it is through them that a textual description is serialized into the token sequence that the receiver's generative model finally consumes.
The size of $\cV$ is $T=32{,}128$, which is also the cardinality of $\cT$.
Even if the transmitter does not necessarily host a T5-XXL model, its tokenizer and detokenizer occupy merely $0.8$~MB of storage, which is readily affordable.

For the packet-erasure channel of Sec.~\ref{ssec_channel_model}, each token sequence is interleaved over $P=10$ packets according to~\eqref{eq_interleave}, and every packet is erased independently with probability $\gamma$.
From the set of channel conditions $\cR=[0.1, 0.7]$, three representative erasure rates are selected as $\gamma_{\mathrm{L}}=0.2$, $\gamma_{\mathrm{M}}=0.4$, and $\gamma_{\mathrm{H}}=0.6$, i.e., $\cR_{\mathrm{r}}=\{0.2,0.4,0.6\}$.
In each realization, at least one randomly chosen packet is ensured to survive because the realization in which all the $P$ packets are lost is uninformative.
At the receiver, the surviving tokens are restored to their original relative order and directly concatenated without any padding tokens at the erased positions.
This is because such tokens are out of the training distribution of the text-to-image generative model and are empirically observed to degrade the image generation quality substantially.

For the target textual descriptions to be transmitted, we draw them from DiffusionDB~\cite{wang2023diffusiondb}, a large-scale collection of prompts written by real users of text-to-image services, which covers diverse subjects, lengths, and writing styles.
Specifically, $3{,}797$ distinct prompts whose token sequence lengths satisfy $N\in[20,100]$ are randomly sampled, and the reference image of each prompt is generated by PixArt-Sigma under a fixed random seed of $42$.
Each prompt is thus a sufficient description of its reference image, and the receiver can reconstruct the image exactly when no token is lost.
The prompts are randomly split into $3{,}038$ for training and $759$ for evaluation.
The training prompts constitute the dataset $\cD_{\mathrm{pmt}}$ on which \alg\ trains the token encoder in \sys.

\subsection{Training Configuration}
\label{ssec_implementation}

The token encoder of \sys is instantiated by Qwen3-8B-Base~\cite{Yang25arXiv_Qwen3}, an open-source LLM.
Its vocabulary, $\widehat{\cV}$, is of size $\widehat{T}=151{,}936$, and thus differs substantially from $\cV$.
The talent model of \alg is instantiated by T5-XXL, which meets the two requirements stated in Sec.~\ref{ssec_stage1_expert_tuning}.
The sentence embedder, $\bm\xi(\cdot)$, is instantiated by Sentence-T5~\cite{Ni22ACL_SentenceT5}, which is associated with the same vocabulary and can therefore take the output of the talent directly.

When training the talent to an expert, the talent is adapted by a low-rank adapter of rank $r=128$, injected into the query and value projections of every block in T5-XXL's decoder module, while its encoder module is kept frozen to preserve its understanding of semantics.
In total, the adapter comprises $101$ million trainable parameters, amounting to $1.58\%$ of the $6.37$ billion parameters of the decoder.
Based on the talent, each expert for $\gamma\in\cR_{\mathrm{r}}$ is formed by optimizing~\eqref{eq_obj_sentence_total} with $\lambda=0.05$ for $3{,}000$ steps at a batch size of $32$ prompts, initialized by the identity-copy warm-up.

The three experts then encode every prompt of $\cD_{\mathrm{pmt}}$ in a deterministic manner, which yields three distillation datasets $\{\cD_{\mathrm{H}},\cD_{\mathrm{M}},\cD_{\mathrm{L}}\}$.
For the student, the reconfigurable low-rank parameterization in~\eqref{eq_gated_lora_update} is injected into the query, key, value, and output projections of every attention block with rank $r=64$ by default.
The reweighting function embeds the scalar erasure rate into a $16$-dimensional feature $\bigl[\sin(2^{k}\pi\gamma),\,\cos(2^{k}\pi\gamma)\bigr]_{k=0}^{7}$, which is then mapped to the $r$ scaling factors through a two-layer perceptron of hidden dimension $16$.
The resulting adapter comprises $61.5$ million trainable parameters, which is $0.75\%$ of the $8.2$ billion parameters of the LLM, and only $195{,}840$ of them, i.e., $0.32\%$ of the adapter, belong to the reweighting functions.
The expert distillation in~\eqref{eq_distill_nll} is optimized for $15{,}000$ steps at a batch size of one prompt, with the three representative erasure rates drawn uniformly and independently at every step.

The student reinforcement starts from the distilled student.
At every step, a group of $G=8$ actions is sampled with the number of generated tokens capped by the length of the source sequence in $\widehat{\cT}$, and the parameters are updated over $5{,}000$ steps with $\varepsilon=0.2$ and $\eta=0.01$.
The reward is evaluated by rendering the received token sequence into an image and comparing it with the reference image of the source prompt, where the DINO-score and the CLIP-score are extracted by DINOv3~\cite{Simeoni26TMLR_DINOv3} and CLIP-L/14~\cite{Radford21ICML_CLIP}.
Moreover, to keep the reward evaluation affordable, the images inside the reward loop are rendered by a distilled few-step variant of the generative model, which produces a $512\times512$ image within four denoising steps.
In the testing, images are generated within eight denoising steps instead.

%% file: sections/7b_experiments_results.tex
\section{Simulation Results}
\label{sec_experiment_results}

In this section, we compare \sys with receiver-side semantic recovery benchmarks under a unified protocol, ablate the contribution of each stage of \alg, and evaluate the impact of the low-rank adapter's rank.

\subsection{Overall Performance of \protect\textnormal{\sys}}
All results use the same evaluation set in which $50$ prompts drawn from the $759$ held-out prompts are paired with $K=1,\ldots,10$ surviving packets under $10$ random channel erasure patterns.
Thus, every method is evaluated on the same $5000$ test cases.
The performance is evaluated over different channel conditions, represented by number of surviving packets $K$ instead of erasure rate $\gamma$. 
As for the evaluation metrics, we evaluate the sentence-level semantic recovery by $S_{\mathrm{sent}}$ in~\eqref{eq_obj_sentence_sent} with Sentence-T5, and the image-level one by the average of the DINO-score and CLIP-score as in~\eqref{eq_obj_token_enc}.
We note that all the images are rendered by PixArt-Sigma at $512\times 512$ within eight denoising steps under a fixed random seed.

\ifSingleColumn
    \renewcommand{\figwidth}{1\textwidth}
\else
    \renewcommand{\figwidth}{1\textwidth}
\fi
\begin{figure*}[t]
  \centering
  \includegraphics[width=\figwidth]{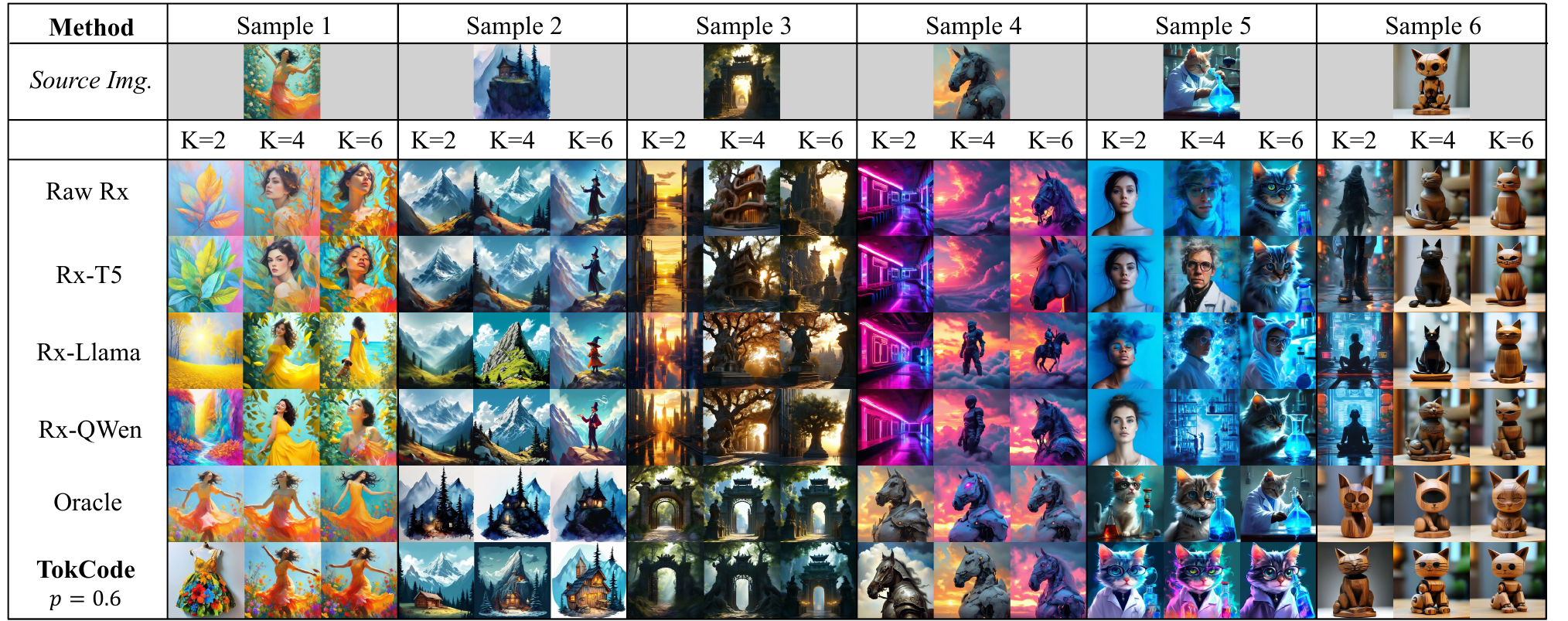}
  \caption{Rx-generated image comparison under $K=2$, $4$, and $6$ surviving packets, where each row corresponds to a method and each column group corresponds to a test sample.}
  \label{fig_exp1}
  \vspace{-1em}
\end{figure*}

To evaluate the advantage of \sys, in addition to comparing it with the baseline method without any recovery, we further compare it with three receiver-side recovery benchmark methods, each of which performs the recovery with a representative, powerful, and recently released local LLM.
We focus on the receiver-side recovery because, to the best of our knowledge, \sys is the first transmitter-side token encoding framework for semantic recovery, and hence no directly comparable benchmark exists.
The four methods are detailed as follows.
\begin{itemize}[leftmargin=*]
\item \emph{Raw Rx}: The source token sequence is transmitted without encoding, and the surviving tokens are directly concatenated without any recovery.

\item \emph{Rx-T5}: Following the receiver-side generative recovery principle in~\cite{Qiao25WCM_Token}, every erased position is infilled by the T5-XXL decoder.
The T5-XXL decoder is chosen since token infilling aligns with its pretraining objective.

\item \emph{Rx-LLaMA}: The complete prompt is predicted from its partially erased version by LLaMA-3-8B-Instruct~\cite{Grattafiori24arXiv_Llama3}.
This LLM is selected as a recently released 8-billion-parameter model that is comparable to Qwen3-8B-Base in size.

\item \emph{Rx-Qwen}: The same recovery is performed by Qwen3-8B-Base, which shares the backbone of the token encoder of \sys and is not trained by \alg, so that the difference between them is attributed to \alg alone.
\end{itemize}

Moreover, we include an \emph{Oracle} scheme as the performance upper bound, which knows in advance which packets will be erased and hence how many source tokens can survive.
Under this budget, the surviving tokens are selected iteratively, where each step adds the token whose inclusion attains the highest Sentence-T5 similarity to the source sentence.
As this selection is greedy and restricted to the source tokens, \emph{Oracle} serves as an approximate rather than an exact upper bound. Compared with the upper-bound approximation in our preliminary version~\cite{Hu26GLOBECOM_TokCode}, where each token is scored individually, this greedy selection better captures the combined semantics of the retained tokens, yielding a stronger approximate upper bound.

Fig.~\ref{fig_exp1} compares receiver-generated images for six samples at $K=2,4,6$.
For each sample and $K$, all methods share the same erasure pattern and generative-model seed, so that differences are attributable to the methods alone.
The surviving packet sets are nested across $K$, making changes across $K$ reflect additional delivered tokens rather than different erasure patterns.
Under this protocol, \emph{Raw Rx} suffers significant semantic distortion on all the samples, and \emph{Rx-T5} barely alleviates it.
\emph{Rx-LLaMA} and \emph{Rx-Qwen} recover part of the target semantics on some samples, e.g., the first and the fourth samples at $K=4$, but the recovered semantics still drift away from the target.
In contrast, \sys preserves the key semantics of the target images in most cases, even at $K=2$, and is visibly the closest to \emph{Lossless} among the non-oracle methods.
Its outputs are comparable to those of \emph{Oracle}, despite the latter being granted the erasure pattern in advance.


\input{sections/table_1}

In Table~\ref{tab_1}, taking Sample~1 at $K=4$ in Fig.~\ref{fig_exp1} as an example, we compare the token sequences fed to the text-to-image model by different methods, which explains the above differences at the text level.
\emph{Raw Rx} loses key concepts, such as ``happy,'' ``colorful,'' ``dress,'' and ``dances,'' and hence the expression and the pose of the lady in its generated image differ significantly from those in the source image.
\emph{Rx-T5} fills the erased positions with short and disordered fragments and restores none of the lost concepts, which explains why its generated image is close to that of \emph{Raw Rx}.
This behavior of \emph{Rx-T5} is attributed to the erased fraction being much more than the masking ratio encountered by the T5-XXL decoder in its pretraining.
\emph{Rx-LLaMA} and \emph{Rx-Qwen} both return fluent prompts, but the recovered semantics are shifted: the dress becomes ``yellow'' rather than ``colorful,'' the dancing action and the jasmines are dropped, and unmentioned details such as a ``smile'' are introduced.
Comparing them with the surviving part of \emph{Raw Rx} shows that, once a concept is erased, the receiver has no evidence from which to recover it.
By knowing the surviving positions in advance, \emph{Oracle} successfully retains the key concepts, e.g., ``happy lady,'' ``colorful flowy dress dances,'' and ``pastel,'' and hence its generated image is closer to the source image than those of the above methods.
In contrast, \sys drops the non-critical tokens and repeats the critical ones at multiple positions, so that ``happy,'' ``lady,'' ``colorful,'' ``dress,'' ``dancing,'' ``flowers,'' ``leaves,'' and ``sunny'' all survive the erasure, and hence its generated image bears high visual similarity to the source image.

\ifSingleColumn
    \renewcommand{\figwidth}{0.75\linewidth}
\else
    \renewcommand{\figwidth}{1\linewidth}
\fi
\begin{figure}[t]
    \centering
    \includegraphics[width=\figwidth]{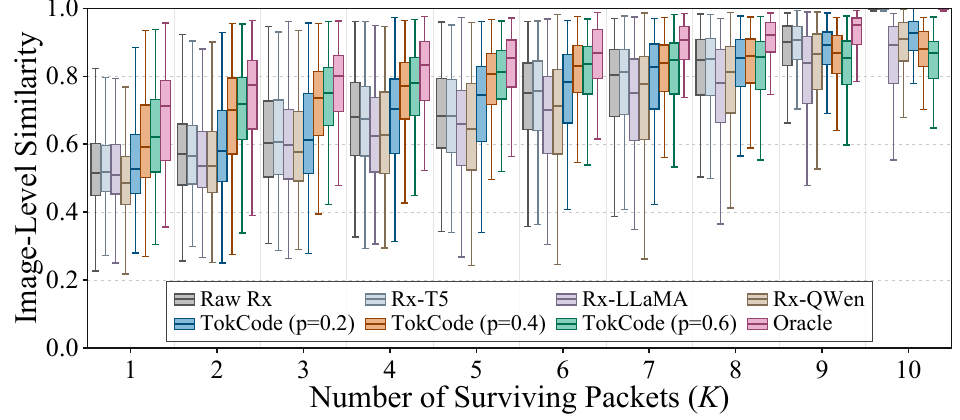}
    \caption{Resulting image-level similarity comparison under different numbers of surviving packets.}
    \label{fig_exp3}
    \vspace{-1em}
\end{figure}

Fig.~\ref{fig_exp3} compares image-level similarity.
Receiver-side recovery barely improves similarity: \emph{Rx-T5} remains close to \emph{Raw Rx} while \emph{Rx-LLaMA} and \emph{Rx-Qwen} fall below it.
The limited gain of receiver-side recovery supports our motivation that, without semantic encoding at the transmitter, losses of critical semantics can hardly be recovered at the receiver.
With reweighting input $p=0.6$, \sys achieves clear gains under severe packet loss ($K\leq6$) compared to all the receiver-side recovery methods and the \emph{Raw Rx} baseline.
Notably, at $K=2$, $3$, $4$, and $5$, it improves the mean image-level similarity over the best-performing benchmark by $22.4\%$, $19.7\%$, $14.1\%$, and $16.3\%$, respectively, closing $72.0\%$, $76.5\%$, $71.9\%$, and $76.4\%$ of the gap between that benchmark and \emph{Oracle}.

When more packets survive, e.g., $K\geq8$, \sys may lead to a lower image-level similarity than \emph{Raw Rx} and \emph{Rx-T5}.
This exposes the essential characteristic of \sys: it trades off the fidelity for the robustness against the channel since the positions taken by the repeated key tokens can no longer carry fine-grained details.
Such a trade-off is favorable in a challenging wireless environment with a high $\gamma$, where $K$ concentrates at low values.
The reweighting input $p=0.2$ or $p=0.4$ narrows this gap at large $K$, at the cost of the performance at small $K$.
Our design in~\eqref{eq_gate_selection} enables an optimal selection of the reweighting input under the distribution of $K$ induced by the actual $\gamma$.

\subsection{Stage-Wise Ablation of \protect\textnormal{\alg}}


\ifSingleColumn
    \renewcommand{\figwidth}{0.65\linewidth}
\else
    \renewcommand{\figwidth}{0.9\linewidth}
\fi
\begin{figure}[t]
    \centering
    \includegraphics[width=\figwidth]{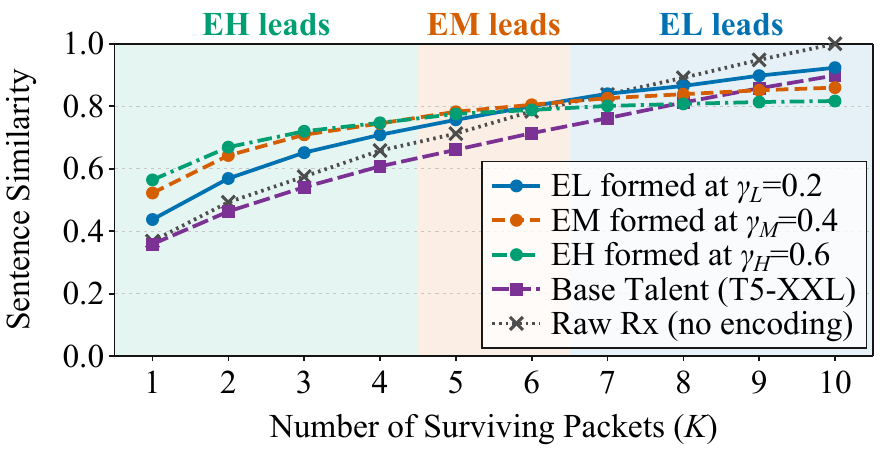}
    \vspace{-1em}
    \caption{Comparison of the semantic recovery performance of experts formed under different packet-erasure rates.}
    \label{fig_exp_stage1}
    \vspace{-1em}
\end{figure}

Fig.~\ref{fig_exp_stage1} compares experts $\bm\zeta_{\mathrm{L}}(\cdot)$, $\bm\zeta_{\mathrm{M}}(\cdot)$, and $\bm\zeta_{\mathrm{H}}(\cdot)$, referred to as \emph{EL}, \emph{EM}, and \emph{EH}, that are formed in the expert formation stage of \alg.
Among the three experts, \emph{EH} leads for $K\leq4$, \emph{EM} for $K=5$ and $6$, and \emph{EL} for $K\geq7$, matching the intuitive expectation that an expert trained under a higher erasure rate suits a smaller number of surviving packets.
That the leading expert changes with $K$ demonstrates that the training under different channel conditions has led each expert to form a semantic encoding strategy that suits its suffered erasure level.
Moreover, every expert also clearly outperforms the base talent within its own range, which verifies the effectiveness of the expert formation stage of \alg.
For $K\geq8$, the \emph{Raw Rx} baseline regains the lead as most of the source tokens survive, which is consistent with the results observed in Fig.~\ref{fig_exp3}.


\ifSingleColumn
    \renewcommand{\figwidth}{0.75\linewidth}
\else
    \renewcommand{\figwidth}{1\linewidth}
\fi
\begin{figure}[t]
    \centering
    \includegraphics[width=\figwidth]{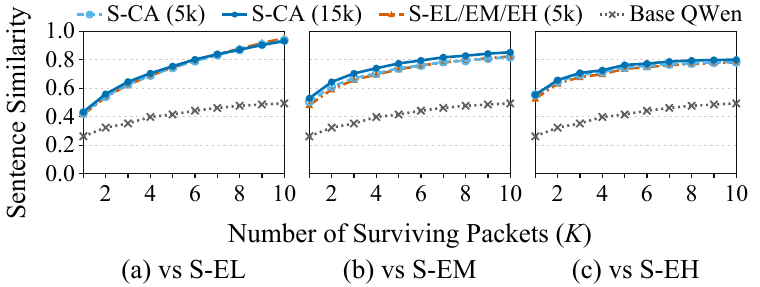}
    \vspace{-2em}
    \caption{Performance comparison among the student with the reconfigurable low-rank adapter and the per-expert students under different numbers of surviving packets.}
    \label{fig_exp_stage2}
    \vspace{-1em}
 \end{figure}

Fig.~\ref{fig_exp_stage2} evaluates the second stage, i.e., the expert distillation, where a single student \emph{S-CA} is distilled jointly from the three experts and compared with the students \emph{S-EL}, \emph{S-EM}, and \emph{S-EH}, each distilled from one expert.
Under the same $5$k distillation steps, \emph{S-CA} performs on par with the three per-expert students, where the average sentence similarity over $K=1,\ldots,10$ differs by only $-0.4\%$, $+0.9\%$, and $+1.5\%$, respectively, showing that the channel-quality-aware distillation does not weaken the adaptation to each channel condition despite the parameter sharing.
Moreover, under a matched total budget, i.e., $15$k distillation steps, \emph{S-CA} further outperforms the three per-expert students by $0.7\%$, $5.6\%$, and $3.6\%$, respectively.
This verifies the effectiveness of realizing the channel-quality-aware distillation by reweighting the rank-one components in~\eqref{eq_gated_lora_update}.


\ifSingleColumn
    \renewcommand{\figwidth}{0.37\linewidth}
\else
    \renewcommand{\figwidth}{0.49\linewidth}
\fi
\begin{figure}[t]
    \centering
    \includegraphics[width=\figwidth]{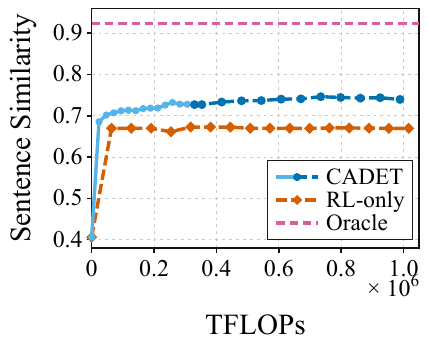}
    \ifSingleColumn
    \else
    \hfill
    \fi
    \includegraphics[width=\figwidth]{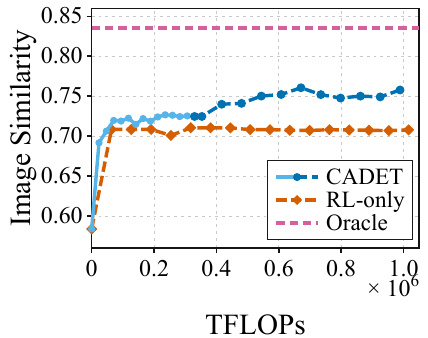}
    \vspace{-2em}
    \caption{Ablation of the expert distillation on the student reinforcement, in terms of the sentence similarity (left) and the image-level similarity (right). The y-axis ranges are tightened to clearly present the performance gaps.} 
    \label{fig_exp_stage3}
    \vspace{-1em}
\end{figure}

Fig.~\ref{fig_exp_stage3} compares \alg with \emph{RL-only}, which devotes the entire student training budget in FLOPs to the reinforcement learning.
Specifically, \alg undergoes $15{,}000$ distillation steps, plotted as the solid lines with dense markers as each such step costs far fewer FLOPs, followed by $5{,}000$ reinforcement steps, plotted as the dashed lines with sparse markers.
\emph{RL-only} spends the same cumulative FLOPs on $7{,}500$ reinforcement steps alone.
\emph{RL-only} improves rapidly at the beginning but then stays at a sentence similarity of around $0.67$ and an image-level similarity of around $0.71$, which confirms the plateau anticipated in Sec.~\ref{ssec_objective_definition}.
In contrast, the expert distillation of \alg directly transfers the encoding strategies of the experts to the student, and thus carries it beyond such a plateau, allowing the reinforcement learning to raise the image-level similarity to $0.761$.
At the budget-matched endpoints, \alg exceeds \emph{RL-only} by $10.5\%$ and $7.2\%$ in the two similarities, closing its gap to \emph{Oracle} by $27.6\%$ and $39.6\%$, respectively.

\subsection{Impact of the Adapter Size}


\ifSingleColumn
    \renewcommand{\figwidth}{0.37\linewidth}
\else
    \renewcommand{\figwidth}{0.49\linewidth}
\fi
\begin{figure}[t]
    \centering
    \includegraphics[width=\figwidth]{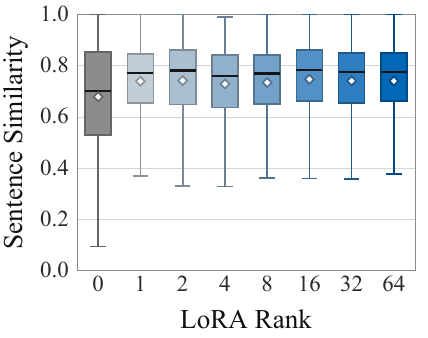}
    \ifSingleColumn
    \else
    \hfill
    \fi
    \includegraphics[width=\figwidth]{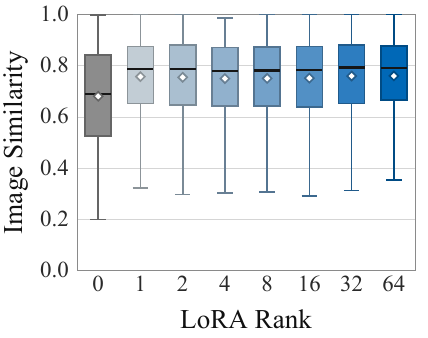}
    \vspace{-2em}
    \caption{Impact of the rank on the semantic recovery performance, in terms of the per-sample sentence similarity (left) and the image-level similarity (right).}
    \label{fig_exp_stage3_rank}
    \vspace{-1em}
\end{figure}

Finally, we examine the impact of the adapter size on the performance of \sys, which directly determines the computation resource and storage overhead required for the token encoding.
The adapter size is proportional to rank $r$ of the low-rank adaptation matrices. 
We sweep $r$ from $64$ down to $1$ and report the resulting performance in Fig.~\ref{fig_exp_stage3_rank}, where each box pools the $50$ evaluation prompts over $K=1,\ldots,10$ and $p\in\{0.2,0.4,0.6\}$, and rank $0$ denotes the transmitter's local LLM without any adapter.
Reducing the rank over this range shrinks the adapter substantially, whereas the training FLOPs decrease only subtly (about $1.0\%$), as the large frozen backbone dominates the training cost.
Moreover, in contrast to the intuition that a smaller rank should degrade the performance, the two similarities stay close across all the tested ranks and fluctuate non-monotonically, which we attribute to the randomness of the training and to the fact that a larger rank offers more capacity but also is harder to optimize.
Consequently, the adapter remains effective even at $r=1$, where its size becomes $98.4\%$ smaller than at $r=64$, occupying only about $2$ MB in the FP16 format.

%% file: sections/table_1.tex
\begingroup
\definecolor{EncodedRed}{HTML}{C62828}
\definecolor{GeneratedBlue}{HTML}{1565C0}
\definecolor{LLMGreen}{HTML}{2E7D32}
\newcommand{\striketok}[1]{\begingroup\setbox0=\hbox{#1}\rlap{\raisebox{0.45ex}{\rule{\wd0}{0.38pt}}}#1\endgroup}
\newcommand{\enctok}[1]{\textcolor{EncodedRed}{#1}}
\newcommand{\gentok}[1]{\textcolor{GeneratedBlue}{#1}}
\newcommand{\llmtok}[1]{\textcolor{LLMGreen}{#1}}
\newcommand{\nontok}[1]{\texttt{<#1>}}
\renewcommand{\arraystretch}{1.10}
\setlength{\tabcolsep}{2pt}
\begin{table}[t]
\centering
\caption{Comparison among inputs to the text-to-image model for the first sample of Fig.~\ref{fig_exp1} under $K=4$.}
\label{tab_1}
{\fontsize{6.2}{7.4}\selectfont
\begin{tabular}{|>{\centering\arraybackslash}m{17mm}|>{\raggedright\arraybackslash}m{68mm}|}
\hline
\textbf{Case} & \textbf{Corresponding text input to the text-to-image model} \\
\hline
Lossless & \nontok{SPACE}\allowbreak{} a\allowbreak{} happy\allowbreak{} lady\allowbreak{} in\allowbreak{} \nontok{SPACE}\allowbreak{} a\allowbreak{} bright\allowbreak{} colorful\allowbreak{} flowy\allowbreak{} dress\allowbreak{} dances\allowbreak{} among\allowbreak{} jasmines\allowbreak{} and\allowbreak{} leaves\allowbreak{} on\allowbreak{} \nontok{SPACE}\allowbreak{} a\allowbreak{} bright\allowbreak{} sunny\allowbreak{} day\allowbreak{} with\allowbreak{} her\allowbreak{} arms\allowbreak{} \nontok{SPACE}\allowbreak{} splayed\allowbreak{} open.\allowbreak{} by\allowbreak{} \nontok{SPACE}\allowbreak{} rebecca\allowbreak{} \nontok{SPACE}\allowbreak{} guay.\allowbreak{} pastel\allowbreak{} tones\allowbreak{} \allowbreak{} \\
\hline
Raw Rx & \nontok{SPACE}\allowbreak{} \striketok{a}\allowbreak{} \striketok{happy}\allowbreak{} lady\allowbreak{} in\allowbreak{} \striketok{\nontok{SPACE}}\allowbreak{} \striketok{a}\allowbreak{} bright\allowbreak{} \striketok{colorful}\allowbreak{} \striketok{flow}y\allowbreak{} \striketok{dress}\allowbreak{} \striketok{dance}s\allowbreak{} among\allowbreak{} \striketok{ja}\striketok{s}mine\striketok{s}\allowbreak{} \striketok{and}\allowbreak{} leaves\allowbreak{} \striketok{on}\allowbreak{} \striketok{\nontok{SPACE}}\allowbreak{} a\allowbreak{} bright\allowbreak{} \striketok{sunny}\allowbreak{} \striketok{day}\allowbreak{} with\allowbreak{} \striketok{her}\allowbreak{} \striketok{arms}\allowbreak{} \nontok{SPACE}\allowbreak{} \striketok{s}\striketok{play}ed\allowbreak{} \striketok{open}\striketok{.}\allowbreak{} by\allowbreak{} \striketok{\nontok{SPACE}}\allowbreak{} \striketok{re}be\striketok{cca}\allowbreak{} \striketok{\nontok{SPACE}}\allowbreak{} gua\striketok{y}\striketok{.}\allowbreak{} pastel\allowbreak{} \striketok{to}\striketok{nes}\allowbreak{} \allowbreak{} \\
\hline
Rx-T5 & \nontok{SPACE}\allowbreak{} \striketok{a}\gentok{a}\allowbreak{} \striketok{happy}\gentok{a}\allowbreak{} lady\allowbreak{} in\allowbreak{} \striketok{\nontok{SPACE}}\allowbreak{} \allowbreak{} \gentok{the}\striketok{a}\allowbreak{} \gentok{the}\allowbreak{} bright\allowbreak{} \striketok{colorful}\gentok{.}\allowbreak{} \striketok{flow}\gentok{.}y\allowbreak{} \striketok{dress}\allowbreak{} \gentok{\nontok{SPACE}}\allowbreak{} \allowbreak{} \striketok{dance}\allowbreak{} \gentok{\nontok{SPACE}}\allowbreak{} s\allowbreak{} among\allowbreak{} \striketok{ja}\allowbreak{} \gentok{ja}\striketok{s}\allowbreak{} \gentok{ja}mine\striketok{s}\gentok{s}\allowbreak{} \striketok{and}\gentok{s}\allowbreak{} leaves\allowbreak{} \striketok{on}\allowbreak{} \gentok{\nontok{SPACE}}\allowbreak{} \allowbreak{} \striketok{\nontok{SPACE}}\allowbreak{} \allowbreak{} \gentok{\nontok{SPACE}}\allowbreak{} a\allowbreak{} bright\allowbreak{} \striketok{sunny}\gentok{y}\allowbreak{} \striketok{day}\gentok{y}\allowbreak{} with\allowbreak{} \striketok{her}\allowbreak{} \gentok{\nontok{SPACE}}\allowbreak{} \allowbreak{} \striketok{arms}\allowbreak{} \gentok{\nontok{SPACE}}\allowbreak{} \allowbreak{} \nontok{SPACE}\allowbreak{} \striketok{s}\gentok{a}\striketok{play}\gentok{a}ed\allowbreak{} \striketok{open}\gentok{g}\striketok{.}\gentok{g}\allowbreak{} by\allowbreak{} \striketok{\nontok{SPACE}}\allowbreak{} \allowbreak{} \gentok{\nontok{SPACE}}\allowbreak{} \striketok{re}\allowbreak{} \gentok{\nontok{SPACE}}\allowbreak{} be\striketok{cca}\gentok{au}\allowbreak{} \striketok{\nontok{SPACE}}\allowbreak{} \gentok{au}gua\striketok{y}\gentok{s}\striketok{.}\gentok{s}\allowbreak{} pastel\allowbreak{} \striketok{to}\gentok{s}\striketok{nes}\gentok{s}\allowbreak{} \allowbreak{} \\
\hline
Rx-LLaMA & \llmtok{Lady}\allowbreak{} \llmtok{in}\allowbreak{} \llmtok{\nontok{SPACE}}\allowbreak{} \llmtok{a}\allowbreak{} \llmtok{bright}\allowbreak{} \llmtok{yellow}\allowbreak{} \llmtok{dress}\allowbreak{} \llmtok{among}\allowbreak{} \llmtok{leaves}\allowbreak{} \llmtok{with}\allowbreak{} \llmtok{\nontok{SPACE}}\allowbreak{} \llmtok{a}\allowbreak{} \llmtok{bright}\allowbreak{} \llmtok{smile}\llmtok{,}\allowbreak{} \llmtok{\nontok{SPACE}}\allowbreak{} \llmtok{surrounded}\allowbreak{} \llmtok{by}\allowbreak{} \llmtok{\nontok{SPACE}}\allowbreak{} \llmtok{a}\allowbreak{} \llmtok{soft}\allowbreak{} \llmtok{pastel}\allowbreak{} \llmtok{glow}\llmtok{.}\allowbreak{} \llmtok{}\allowbreak{} \\
\hline
Rx-Qwen & \llmtok{A}\allowbreak{} \llmtok{lady}\allowbreak{} \llmtok{in}\allowbreak{} \llmtok{\nontok{SPACE}}\allowbreak{} \llmtok{a}\allowbreak{} \llmtok{bright}\allowbreak{} \llmtok{yellow}\allowbreak{} \llmtok{dress}\allowbreak{} \llmtok{among}\allowbreak{} \llmtok{mine}\allowbreak{} \llmtok{leaves}\llmtok{,}\allowbreak{} \llmtok{\nontok{SPACE}}\allowbreak{} \llmtok{a}\allowbreak{} \llmtok{bright}\allowbreak{} \llmtok{smile}\allowbreak{} \llmtok{with}\allowbreak{} \llmtok{\nontok{SPACE}}\allowbreak{} \llmtok{a}\allowbreak{} \llmtok{pastel}\allowbreak{} \llmtok{backdrop}\llmtok{.}\allowbreak{} \llmtok{}\allowbreak{} \\
\hline
Oracle & \nontok{SPACE}\allowbreak{} a\allowbreak{} happy\allowbreak{} lady\allowbreak{} \striketok{in}\allowbreak{} \nontok{SPACE}\allowbreak{} \striketok{a}\allowbreak{} \striketok{bright}\allowbreak{} colorful\allowbreak{} flowy\allowbreak{} dress\allowbreak{} dances\allowbreak{} among\allowbreak{} \striketok{ja}\striketok{s}\striketok{mine}\striketok{s}\allowbreak{} \striketok{and}\allowbreak{} \striketok{leaves}\allowbreak{} \striketok{on}\allowbreak{} \striketok{\nontok{SPACE}}\allowbreak{} \striketok{a}\allowbreak{} \striketok{bright}\allowbreak{} sunny\allowbreak{} \striketok{day}\allowbreak{} \striketok{with}\allowbreak{} \striketok{her}\allowbreak{} \striketok{arms}\allowbreak{} \striketok{\nontok{SPACE}}\allowbreak{} \striketok{s}\striketok{play}\striketok{e}\striketok{d}\allowbreak{} open\striketok{.}\allowbreak{} \striketok{by}\allowbreak{} \striketok{\nontok{SPACE}}\allowbreak{} rebecca\allowbreak{} \nontok{SPACE}\allowbreak{} gu\striketok{a}\striketok{y}\striketok{.}\allowbreak{} pastel\allowbreak{} \striketok{to}\striketok{nes}\allowbreak{} \allowbreak{} \\
\hline
\textbf{\sys ($p=0.6$)} & \enctok{happy}\allowbreak{} \enctok{\striketok{happy}}\allowbreak{} \striketok{happy}\allowbreak{} \enctok{happy}\allowbreak{} \enctok{lady}\allowbreak{} \enctok{\striketok{lady}}\allowbreak{} \enctok{\striketok{lady}}\allowbreak{} \enctok{happy}\allowbreak{} \enctok{\striketok{lady}}\allowbreak{} \enctok{\striketok{happy}}\allowbreak{} \enctok{happy}\allowbreak{} \enctok{\striketok{colorful}}\allowbreak{} \enctok{\striketok{colorful}}\allowbreak{} \enctok{happy}\allowbreak{} \enctok{dress}\allowbreak{} \enctok{\striketok{colorful}}\allowbreak{} \enctok{\striketok{colorful}}\allowbreak{} \enctok{colorful}\allowbreak{} \enctok{\striketok{dress}}\allowbreak{} \enctok{\striketok{dress}}\allowbreak{} \enctok{dress}\allowbreak{} \enctok{\striketok{dancing}}\allowbreak{} \enctok{\striketok{dancing}}\allowbreak{} \enctok{dancing}\allowbreak{} \enctok{flowers}\allowbreak{} \enctok{\striketok{ja}}\enctok{\striketok{s}}\enctok{mine}\allowbreak{} \enctok{\striketok{ja}}\enctok{\striketok{s}}\enctok{mine}\allowbreak{} \enctok{\striketok{ja}}\enctok{\striketok{s}}\enctok{mine}\allowbreak{} \enctok{ja}\enctok{\striketok{s}}\enctok{\striketok{mine}}\allowbreak{} \enctok{flowers}\allowbreak{} \enctok{\striketok{flowers}}\allowbreak{} \enctok{\striketok{flowers}}\allowbreak{} \enctok{leaves}\allowbreak{} \enctok{\striketok{leaves}}\allowbreak{} \enctok{\striketok{sunny}}\allowbreak{} \enctok{sunny}\allowbreak{} \enctok{sunny}\allowbreak{} \enctok{\striketok{happy}}\allowbreak{} \enctok{\striketok{lady}}\allowbreak{} \allowbreak{} \\
\hline
\end{tabular}
}
\par\vspace{5pt}
\noindent\parbox{\columnwidth}{\raggedright\fontsize{6.8}{8.0}\selectfont
\emph{Note}: Every token is displayed by its corresponding text piece in the vocabulary.
Black: unchanged/source token;
Strike (\striketok{ABC}): token erased by the channel;
\enctok{Red}: encoded tokens by \sys;
\gentok{Blue}: recovered token of Rx-T5;
\llmtok{Green}: predicted prompt at the receiver.}
\vspace{-0.8em}
\end{table}
\endgroup

%% file: sections/9_conclusion.tex
\section{Conclusion and Remarks}
\label{sec_conclusion}

In this paper, we have proposed a token encoding framework, \sys, for robust semantic recovery in generative semantic communication systems in challenging wireless environments.
\sys recasts a general-purpose local LLM at the transmitter into a token encoder through a lightweight adapter, restructuring redundancy in the semantic domain without any extra vocabulary, extra tokens, or receiver-side restoration.
To train the adapter, we have developed \alg, which tunes text foundation models into erasure-rate-specific experts under a differentiable sentence-level semantic surrogate, distills them into a single reconfigurable low-rank adapter, and then refines it by reinforcement learning.
Simulation results on token-based generative image transmission show that, when only $20\%$ to $50\%$ of the tokens survive, \sys improves the image-level similarity over the best-performing receiver-side recovery benchmark by $14.1\%$--$22.4\%$, closing $71.9\%$--$76.5\%$ of its gap to the erasure-aware oracle encoding.
Moreover, \alg carries the student beyond the plateau at which direct reinforcement learning saturates, closing $39.6\%$ of its gap to the oracle encoding, and retains such an advantage with an adapter of merely $2$~MB.

The proposed \sys framework can be extended to textual-token-based generative semantic communication systems for text, speech, video, and other modalities. For these extensions, the first two stages of \alg retain their procedures while their training data are replaced with real-world prompt datasets for text-to-text/speech/video generation. The student reinforcement stage can be adapted by replacing the image-level reward with a similarity measure aligned with the target modality. Future work will evaluate these extensions and demonstrate the broader applicability of \sys for enhancing the robustness of multi-modal generative semantic communication in challenging wireless environments.